\documentclass[useAMS,fleqn]{mn2e}

\usepackage[dvips]{graphicx}
\usepackage{amssymb}
\usepackage{amsmath}
\usepackage{multirow}
\usepackage{fnpara}
\usepackage{savefnmark}
\def\jnl@style{\rm}
\def\aaref@jnl#1{{\jnl@style#1}}

\def\aaref@jnl#1{{\jnl@style#1\thinspace}}

\def\aj{\aaref@jnl{AJ}}                   
\def\araa{\aaref@jnl{ARA\&A}}             
\def\apj{\aaref@jnl{ApJ}}                 
\def\apjl{\aaref@jnl{ApJ}}                
\def\apjs{\aaref@jnl{ApJS}}               
\def\ao{\aaref@jnl{Appl.~Opt.}}           
\def\apss{\aaref@jnl{Ap\&SS}}             
\def\aap{\aaref@jnl{A\&A}}                
\def\aapr{\aaref@jnl{A\&A~Rev.}}          
\def\aaps{\aaref@jnl{A\&AS}}              
\def\azh{\aaref@jnl{AZh}}                 
\def\baas{\aaref@jnl{BAAS}}               
\def\jrasc{\aaref@jnl{JRASC}}             
\def\memras{\aaref@jnl{MmRAS}}            
\def\mnras{\aaref@jnl{MNRAS}}             
\def\pra{\aaref@jnl{Phys.~Rev.~A}}        
\def\prb{\aaref@jnl{Phys.~Rev.~B}}        
\def\prc{\aaref@jnl{Phys.~Rev.~C}}        
\def\prd{\aaref@jnl{Phys.~Rev.~D}}        
\def\pre{\aaref@jnl{Phys.~Rev.~E}}        
\def\prl{\aaref@jnl{Phys.~Rev.~Lett.}}    
\def\pasp{\aaref@jnl{PASP}}               
\def\pasj{\aaref@jnl{PASJ}}               
\def\qjras{\aaref@jnl{QJRAS}}             
\def\skytel{\aaref@jnl{S\&T}}             
\def\solphys{\aaref@jnl{Sol.~Phys.}}      
\def\sovast{\aaref@jnl{Soviet~Ast.}}      
\def\ssr{\aaref@jnl{Space~Sci.~Rev.}}     
\def\zap{\aaref@jnl{ZAp}}                 
\def\nat{\aaref@jnl{Nature}}              
\def\iaucirc{\aaref@jnl{IAU~Circ.}}       
\def\aplett{\aaref@jnl{Astrophys.~Lett.}} 
\def\apspr{\aaref@jnl{Astrophys.~Space~Phys.~Res.}}
\def\bain{\aaref@jnl{Bull.~Astron.~Inst.~Netherlands}} 
\def\fcp{\aaref@jnl{Fund.~Cosmic~Phys.}}  
\def\gca{\aaref@jnl{Geochim.~Cosmochim.~Acta}}   
\def\grl{\aaref@jnl{Geophys.~Res.~Lett.}} 
\def\jcp{\aaref@jnl{J.~Chem.~Phys.}}      
\def\jgr{\aaref@jnl{J.~Geophys.~Res.}}    
\def\jqsrt{\aaref@jnl{J.~Quant.~Spec.~Radiat.~Transf.}}
\def\memsai{\aaref@jnl{Mem.~Soc.~Astron.~Italiana}}
\def\nphysa{\aaref@jnl{Nucl.~Phys.~A}}   
\def\physrep{\aaref@jnl{Phys.~Rep.}}   
\def\physscr{\aaref@jnl{Phys.~Scr}}   
\def\planss{\aaref@jnl{Planet.~Space~Sci.}}   
\def\procspie{\aaref@jnl{Proc.~SPIE}}   

\begin{document}

\title[Comprehensive simulations of superhumps]
  {Comprehensive simulations of superhumps}
\author[A. J. Smith et al.]
  {Amanda J. Smith$^{1,}\thanks{E-mail: amanda.smith@open.ac.uk}$, Carole A. Haswell$^{1}$, James R. Murray$^{2}$, \and Michael R. Truss$^{3}$ and Stephen B. Foulkes$^{1}$ \\
$^1$Department of Physics \& Astronomy, The Open University, Walton Hall, Milton Keynes, MK7 6AA, UK\\
$^2$Department of Astrophysics \& and Supercomputing, Swinburne University of Technology, Hawthorn, VIC 3122, Australia\\
$^3$Department of Physics, Durham University, South Road, Durham, DH1 3LE, UK}
\date{}
\pagerange{\pageref{firstpage}--\pageref{lastpage}}
\pubyear{2005}
\maketitle \label{firstpage}
\begin{abstract}
We use 3D SPH calculations with higher resolution, as well as with more realistic viscosity and sound-speed prescriptions than previous work to examine the eccentric instability which underlies the superhump phenomenon in semi-detached binaries. We illustrate the importance of the two-armed spiral mode in the generation of superhumps. Differential motions in the fluid disc cause converging flows which lead to strong spiral shocks once each superhump cycle. The dissipation associated with these shocks powers the superhump. We compare 2D and 3D results, and conclude that 3D simulations are necessary to faithfully simulate the disc dynamics. We ran our simulations for unprecedented durations, so that an eccentric equilibrium is established except at high mass ratios where the growth rate of the instability is very low. 

We collate the observed data on superhumps. Our improved simulations give a closer match to the observed relationship between superhump period excess and binary mass ratio than previous numerical work. The observed black hole X-ray transient superhumpers appear to have systematically lower disc precession rates than the cataclysmic variables. This could be due to higher disc temperatures and thicknesses. No high-resolution 3D disc with mass ratio $q > 0.24$ developed superhumps, in agreement with analytical expectations. 

The modulation in total viscous dissipation on the superhump period is overwhelmingly from the region of the disc within the $3:1$ resonance radius. The precession rates of our high resolution 3D discs match the single particle dynamical precession rate at $0.87 R_{3:1}$. As the eccentric instability develops, the viscous torques are enhanced, and the disc consequently adjusts to a new equilibrium state, as suggested in the thermal-tidal instability model. We quantify this enhancement in the viscosity, which is $\sim 10$ per cent for $q = 0.08$. The disc motions can be described as superpositions of the S(k,l) modes, and the disc executes complex standing wave dynamics which repeat in the inertial frame on the disc precession period. We characterise the eccentricity distributions in our accretion discs, and show that the entire body of the disc partakes in the eccentricity. 
\end{abstract}


\section{Introduction}

Cataclysmic variables (CVs) are semi-detached binaries with a Roche-lobe filling low-mass donor star, mass $M_2$, transferring matter onto a white dwarf (WD) primary, mass $M_1$, via an accretion disc. The SU\,UMa-type dwarf novae (DNe) are short-period cataclysmic variables which display two distinct modes of outburst. The normal outbursts are attributed to a thermal--viscous limit--cycle between low and high viscosity states \nocite{O74,H79,MM-H81}({Osaki} 1974; {H{\= o}shi} 1979; {Meyer} \& {Meyer-Hofmeister} 1981), and consequently between low and high mass transfer states \nocite{L01}(see {Lasota} 2001, for a review). The larger amplitude and longer-lasting superoutbursts are characterised by a periodic photometric modulation known as superhumps. Superhumps are attributed to an eccentric apsidally-precessing accretion disc \nocite{V82}({Vogt} 1982). The superhump period, ${P_{\rm sh}}$, is a few percent longer than the orbital period, ${P_{\rm orb}}$: the orientation of the mass donor star relative to the progradely precessing eccentric disc repeats on ${P_{\rm sh}}$. \nocite{Wh88}{Whitehurst} (1988) and \nocite{L91}{Lubow} (1991a) explained that a disc which encounters a 3:1 eccentric inner Lindblad resonance with the tidal potential of the secondary star may become eccentric and precess. A mass ratio $q=M_{2}/M_{1} \lesssim 1/4$ is required for the tidal truncation radius, $R_{\rm tides}$, to lie outside the 3:1 resonance radius, $R_{3:1}$ \nocite{Pac77}({Paczynski} 1977). The thermal--tidal instability (TTI) model \nocite{O89}({Osaki} 1989) attributes the increased brightness and duration of superoutbursts over normal outbursts to an enhanced viscous torque acting once the disc becomes eccentric.

Superhumps arise in several guises, summarised by \nocite{PMR02}{Patterson} {et~al.} (2002a). `Common' or `normal' superhumps ensue after the onset of superoutburst in SU\,UMa systems. They are powered by the periodically varying tidal interaction which modulates dissipation in the disc \nocite{FHM04}(e.g. {Foulkes} {et~al.} 2004). `Late' superhumps sometimes follow common superhumps and are roughly anti-phased to them. The modulation in energy dissipation at accretion stream's impact on the non-axisymmetric disc powers late superhumps \nocite{RHP01}({Rolfe}, {Haswell} \& {Patterson} 2001). `Negative' superhumps, with $P_{\rm sh}$ slightly less than $P_{\rm orb}$, are sometimes observed simultaneously with the more usual `positive' superhumps, and may be related to retrograde precession of a warped accretion disc \nocite{PTS93,MA98,FHM06}({Patterson} {et~al.} 1993b; {Murray} \& {Armitage} 1998; {Foulkes}, {Haswell} \& {Murray} 2006). `Permanent' superhumps are seen in high mass transfer rate systems: the nova-like variables, old novae and some AM\,CVn systems. Some low mass X-ray binaries (LMXBs, the neutron star and black hole analogues of CVs) also show superhumps \nocite{B92,H96,O'DC96}({Bailyn} 1992; {Haswell} 1996; {O'Donoghue} \& {Charles} 1996). In LMXBs optical emission arises overwhelmingly from the reprocessing of X-rays, and superhumps arise from a modulation in reprocessing caused by the changing solid angle subtended by the tidally flexing eccentric disc \nocite{HKM01}({Haswell} {et~al.} 2001). Recently superhumps were reported in the microquasar GRS\,1915+105, which has
an orbital period exceeding 30 days \nocite{NBC06}({Neil}, {Bailyn} \& {Cobb} 2006).

In all the above cases, the fractional superhump excess, $\epsilon=(P_{\rm sh}-P_{\rm orb})/P_{\rm orb}$ varies with $q$, with $|\epsilon |$ increasing with higher values of $q$. The exact relationship has proved difficult to determine. We have performed unprecedentedly comprehensive numerical simulations of apsidally precessing accretion discs, and we present them in the context of previous numerical work, the observational data and salient analytical theory. Section~2 describes our simulations and examines the growth rates of the eccentric instability; the enhanced viscous torques; the superhump light curves which result from the eccentric instability; and compares 2D and 3D simulation results. In section~3 we use two methods to quantify the eccentricity distributions in our simulated discs. Section~4 focuses on the $\epsilon - q$ relationship. In section~5 we discuss our findings. Section~6 gives a summary list of our principle conclusions.

\section{SPH simulations of a Precessing Accretion Disc}
SPH is a Lagrangian method which models fluid flow as a set of moving particles. A detailed review is given by \nocite{Mon92}{Monaghan} (1992). SPH simulations by \nocite{Mu98}{Murray} (1998) provide considerable support for the TTI model, showing that the energy released from a disc that has become tidally unstable is sufficient to account for the excess luminosity of a superoutburst. \nocite{FHM04}{Foulkes} {et~al.} (2004) carried out 2D SPH simulations of a binary system with mass ratio 0.1. They show an eccentric, non-axisymmetric precessing disc of changing density, which is continuously flexing and relaxing on the superhump period. Very clear too in the surface density maps are tightly wrapped spiral density waves which extend from the outermost regions to small radii. They produce shear and dissipation in the outer disc, and propagate angular momentum outwards, allowing disc gas to move inward.

\subsection{Simulation details}
For the calculations presented here, an SPH code is used which has been designed specifically for accretion disc problems. Detailed description of the code can be found in \nocite{Mu96,Mu98}{Murray} (1996, 1998). The code is normalised in units of $a$, the binary separation, for distance; $M_{\rm t} = M_{1}+M_{2}$ for mass; and $P_{\rm orb}/{(2\pi)}$ for time. This code has since been updated to include adaptive spatial resolution, allowing the SPH smoothing length $\lambda$ to vary in both space and time \nocite{MdKL99}({Murray}, {de Kool} \& {Li} 1999). Here, $\lambda$ is set to a maximum value of $0.005\,a$. The code has also been extended to three dimensions \nocite{MA98}({Murray} \& {Armitage} 1998).

Simulations were run for a range of mass ratios, detailed in Table~\ref{tb:sph}.
\begin{table*}
\caption{Summary of simulations and results. The first column denotes the run number. Columns 2 to 6 describe the simulation parameters, recording respectively whether the simulation is conducted in 2D or 3D, the mass ratio, the injection time step, the constant describing the sound speed ($c_{\rm s}={\rm C}\,r^{-3/8}$) and the radius of the primary (the central hole). The time at which the simulation was terminated is given in column 7. The remaining columns record outcomes of the simulations: the mean superhump excess as measured from the simulated lightcurve, the final total number of particles in the simulation, the final average number of neighbours, the strengths of the eccentric and 2-armed spiral modes averaged over the final 10 superhump periods in each simulation, and, as a measure of when the disc initially encounters the resonance, the time at which $S_{(1,0)}=0.01$. $\zeta=1.0$ in all cases.}
\label{tb:sph}
\begin{center}
\begin{minipage}{\textwidth}
\setcounter{mpfootnote}{\value{footnote}}
\renewcommand{\thempfootnote}{\arabic{mpfootnote}}
\begin{tabular}{lllllllllllll}

\hline
Run& 2D/ & $q$ & $\Delta t$ & C & $R_{1}$ & $t_{\rm end}$ & mean $\epsilon$ & np & nne & $S_{(1,0)}$ & $S_{(2,2)}$ & $t_{\rm ecc}$\\& 3D & & ($\Omega_{\rm orb}^{-1}$) & ($a\,\Omega_{\rm orb}$) & ($a$) & ($\Omega_{\rm orb}^{-1}$) & & & & & & ($\Omega_{\rm orb}^{-1}$)\\
\hline
\\
1& 3D & 0.3333 & 0.0025 & 0.050 & 0.03 & 3798.0575 & --- & 123995 & 28.28 & 0.001 & 0.104 & --- \\
2& 3D & 0.2422 & 0.0025 & 0.050 & 0.03 & 8616.6650 & --- & 141474 & 28.50 & 0.001 & 0.099 & --- \\
3& 3D & 0.2346 & 0.0025 & 0.050 & 0.03 & 15496.7600 \footnote{equilibrium not yet reached, simulation continues to run}\saveFN\mass & 0.062 & 143274 & 28.48 & 0.007 & 0.099 & --- \\
4& 3D & 0.2270 & 0.0025 & 0.050 & 0.03 & 14340.9425 \useFN\mass & 0.058 & 143088 & 28.67 & 0.065 & 0.0948 & 6089.7625 \\
5& 3D & 0.2195 & 0.0025 & 0.050 & 0.03 & 15487.8575 & 0.056 & 143730 & 28.18 & 0.092 & 0.092 & 4542.1675 \\
6& 3D & 0.2121 & 0.0025 & 0.050 & 0.03 & 8738.4900 & 0.054 & 144674 & 28.62 & 0.107 & 0.090 & 3188.7950\\
7& 3D & 0.1765 & 0.0025 & 0.050 & 0.03 & 4590.9525 & 0.043 & 150628 & 27.41 & 0.151 & 0.081 & 1271.1125\\
8& 3D & 0.1429 & 0.0025 & 0.050 & 0.03 & 4173.9300 & 0.034 & 160459 & 27.63 & 0.161 & 0.078 & 777.2475\\
9& 3D & 0.1111 & 0.0025 & 0.050 & 0.03 & 3134.1350 & 0.024 & 171971 & 26.92 & 0.172 & 0.075 & 770.7275\\
10& 3D & 0.0811 & 0.0025 & 0.050 & 0.03 & 4323.1750 & 0.017 & 175989 & 24.76 & 0.227 & 0.062 & 477.1300\\
11& 3D & 0.0526 & 0.0025 & 0.050 & 0.03 & 3535.2575 & 0.015 \footnote{temporary appearance of superhumps} & 227047 & 26.76 & 0.001 & 0.080 & 426.7250\\
12& 3D & 0.0256 & 0.0025 & 0.050 & 0.03 & 3491.8375 & --- & 257057 & 24.07 & 0.001 & 0.070 & ---\\
13& 3D & 0.2400 & 0.0025 & 0.050 & 0.05 & 2000.0000 & 0.05 & 118947 & 21.87 & 0.001 & & ---\\
14& 3D & 0.1900 & 0.0025 & 0.050 & 0.05 & 1204.2430 & 0.04 & 128135 & 21.56 & 0.003 & & ---\\
15& 3D & 0.1500 & 0.0025 & 0.050 & 0.05 & 2000.0000 & 0.04 & 141277 & 22.09 & 0.074 & & 1028.375\\
16& 3D & 0.1000 & 0.0025 & 0.050 & 0.05 & 2135.4880 & 0.03 & 145893 & 20.1 & 0.186 & & 622.850\\
17& 3D & 0.0300 & 0.0025 & 0.050 & 0.10 & 942.0200 & --- & 127310 & 27.06 & 0.001 & & ---\\
18& 3D & 0.0700 & 0.0025 & 0.045 & 0.05 & 1014.2300 & 0.01 & 431782 & 98.03 & 0.008 & & ---\\
19& 2D & 0.5385 & 0.0100 & 0.050 & 0.03 & 3081.36 & --- & 18148 & 14.95 & 0.008 & 0.088 & ---\\
20& 2D & 0.4815 & 0.0100 & 0.050 & 0.03 & 21798.61 & 0.093 & 17973 & 14.41 & 0.23 & 0.057 & 1858.91\\
21& 2D & 0.4286 & 0.0100 & 0.050 & 0.03 & 18638.44 & 0.079 & 18293 & 14.35 & 0.30 & 0.043 & 974.19\\
22& 2D & 0.3333 & 0.0100 & 0.050 & 0.03 & 18285.39 & 0.055 & 18493 & 13.85 & 0.43 & 0.020 & 764.33\\
23& 2D & 0.2500 & 0.0100 & 0.050 & 0.03 & 18660.42 & 0.037 & 18822 & 14.89 & 0.53 & 0.019 & 591.20\\
24& 2D & 0.1765 & 0.0100 & 0.050 & 0.03 & 17355.41 & 0.026 & 20438 & 13.27 & 0.56 & 0.021 & 400.43\\
25& 2D & 0.1111 & 0.0100 & 0.050 & 0.03 & 14731.33 & 0.016 & 24766 & 13.97 & 0.54 & 0.020 & 308.67\\
26& 2D & 0.0526 & 0.0100 & 0.050 & 0.03 & 8872.88 & --- & 41753 & 16.00 & 0.001 & 0.081 & ---\\
27& 2D & 0.0256 & 0.0100 & 0.050 & 0.03 & 1608.00 & --- & 46093 & 15.47 & 0.001 & 0.077 & ---\\
[6pt]

\hline
\end{tabular}
\setcounter{footnote}{\value{mpfootnote}}
\end{minipage}
\end{center}
\end{table*}
The simulations were built up from zero mass and a single particle was injected at the $L_{1}$ point each timestep $\Delta t$, so simulating the mass-transfer stream from the secondary. We ran simulations at different mass resolutions, where $\Delta t$ is between $0.01\,\Omega_{\rm orb}^{-1}$ and $0.0025\,\Omega_{\rm orb}^{-1}$. This latter value is at resolution higher than previous calculations \nocite{Mu96,Mu98,FHM04}({Murray} 1996, 1998; {Foulkes} {et~al.} 2004), and we find our results stable to a further reduction in $\Delta t$. The number of particles in the accretion disc in these high-resolution simulations is of order 100\,000, and the average number of `neighbours', i.e. the average number of particles found within $2\lambda$ of each other that are used in the SPH update equations, is between 20 and 30. 

The stream boundary conditions at $L_1$ are a function of $q$, calculated by \nocite{LS75}{Lubow} \& {Shu} (1975) from perturbation analysis of $L_1$. We follow their calculations to determine the direction prograde of the binary axis at which particles are to be injected, $\theta_{\rm inj}$. The initial speed of the particles, $v_{\rm inj}=0.1\,a\Omega_{\rm orb}$, is determined from $c_{\rm s}$ at $L_1$, where the $z$-velocity is an arbitrary fraction of this. The actual value is not critical as the gas will rapidly accelerate to become supersonic, and the stream is not well-resolved in our simulations. For some simulations the third dimension is suppressed. For the inner boundary condition a hole of radius $R_{1}$ centred on the primary was used, with $R_{1}$ set to values between $0.03\,a$ and $0.1\,a$, and particles entering this were removed from the simulation. 

The calculations presented in \nocite{Mu98}{Murray} (1998) and \nocite{MWW00}{Murray}, {Warner} \&  {Wickramasinghe} (2000) were of cool isothermal discs. Here we model a more realistic steady-state disc where $c_{\rm s}$ is a function of disc radius, $r$, and is given by $c_{\rm s}={\rm C}\,r^{-3/8}$. ${\rm C}$ is a constant and in general we have set it equal to $0.05\,a\,\Omega_{\rm orb}$. This means that $c_{\rm s}$ at the resonance radius in each of our simulated discs will be $\simeq 0.067\,a\,\Omega_{\rm orb}$ as detailed in Table~\ref{tb:discs}.
\begin{table*}
\caption{Characterisation of the simulated accretions discs at the 3:1 resonance radius. The 3:1 resonance radius is recorded in the fourth column, followed by the sound speed, the scale height, the characteristic value of the ratio of the disc semi-thickness to the radius as given in Goodchild \& Ogilvie (2006), the shear viscosity, the bulk viscosity, the Shakura--Sunyaev parameter for the shear viscosity and that for the bulk viscosity.}
\label{tb:discs}
\begin{center}
\begin{tabular}{llllllllllllll}

Run & 2D/ & $q$ & $R_{3:1}$ & $c_{\rm s}(R_{3:1})$ & $H(R_{3:1})$ & $h(R_{3:1})$ & $\nu_{\rm sh}(R_{3:1})$ & $\nu_{\rm bk}(R_{3:1})$ & $\alpha_{\rm sh}(R_{3:1})$ & $\alpha_{\rm bk}(R_{3:1})$\\
  & 3D &   & $(a)$ & $(a\Omega_{\rm orb})$ & $(a)$ & $(a)$ & $(a^{2}\Omega_{\rm orb})$ & $(a^{2}\Omega_{\rm orb})$ &   &  \\
\hline
\\
      1 & 3D & 0.3333 & 0.437 & 0.068 & 0.023 & 0.039 & $ 1.55\times10^{-4} $ & $ 3.10\times10^{-4} $ & 0.100 & 0.200\\
      2 & 3D & 0.2422 & 0.447 & 0.068 & 0.023 & 0.038 & $ 1.52\times10^{-4} $ & $ 3.05\times10^{-4} $ & 0.100 & 0.200\\
      3 & 3D & 0.2346 & 0.448 & 0.068 & 0.023 & 0.037 & $ 1.52\times10^{-4} $ & $ 3.04\times10^{-4} $ & 0.100 & 0.200\\
      4 & 3D & 0.2270 & 0.449 & 0.068 & 0.023 & 0.037 & $ 1.52\times10^{-4} $ & $ 3.04\times10^{-4} $ & 0.100 & 0.200\\
      5 & 3D & 0.2195 & 0.450 & 0.067 & 0.022 & 0.037 & $ 1.52\times10^{-4} $ & $ 3.03\times10^{-4} $ & 0.100 & 0.200\\
      6 & 3D & 0.2121 & 0.451 & 0.067 & 0.022 & 0.037 & $ 1.51\times10^{-4} $ & $ 3.03\times10^{-4} $ & 0.100 & 0.200\\
      7 & 3D & 0.1765 & 0.455 & 0.067 & 0.022 & 0.037 & $ 1.50\times10^{-4} $ & $ 3.01\times10^{-4} $ & 0.100 & 0.200\\
      8 & 3D & 0.1429 & 0.460 & 0.067 & 0.022 & 0.036 & $ 1.49\times10^{-4} $ & $ 2.98\times10^{-4} $ & 0.100 & 0.200\\
      9 & 3D & 0.1111 & 0.464 & 0.067 & 0.022 & 0.036 & $ 1.48\times10^{-4} $ & $ 2.96\times10^{-4} $ & 0.100 & 0.200\\
     10 & 3D & 0.0811 & 0.468 & 0.066 & 0.022 & 0.035 & $ 1.47\times10^{-4} $ & $ 2.94\times10^{-4} $ & 0.100 & 0.200\\
     11 & 3D & 0.0526 & 0.473 & 0.066 & 0.022 & 0.035 & $ 1.46\times10^{-4} $ & $ 2.92\times10^{-4} $ & 0.100 & 0.200\\
     12 & 3D & 0.0256 & 0.477 & 0.066 & 0.022 & 0.034 & $ 1.45\times10^{-4} $ & $ 2.91\times10^{-4} $ & 0.100 & 0.200\\
     13 & 3D & 0.2400 & 0.447 & 0.068 & 0.023 & 0.041 & $ 1.52\times10^{-4} $ & $ 3.05\times10^{-4} $ & 0.100 & 0.200\\
     14 & 3D & 0.1900 & 0.454 & 0.067 & 0.022 & 0.040 & $ 1.51\times10^{-4} $ & $ 3.02\times10^{-4} $ & 0.100 & 0.200\\
     15 & 3D & 0.1500 & 0.459 & 0.067 & 0.022 & 0.039 & $ 1.49\times10^{-4} $ & $ 2.99\times10^{-4} $ & 0.100 & 0.200\\
     16 & 3D & 0.1000 & 0.466 & 0.067 & 0.022 & 0.038 & $ 1.48\times10^{-4} $ & $ 2.96\times10^{-4} $ & 0.100 & 0.200\\
     17 & 3D & 0.0300 & 0.476 & 0.066 & 0.022 & 0.037 & $ 1.45\times10^{-4} $ & $ 2.91\times10^{-4} $ & 0.100 & 0.200\\
     18 & 3D & 0.0700 & 0.470 & 0.060 & 0.020 & 0.034 & $ 1.19\times10^{-4} $ & $ 2.38\times10^{-4} $ & 0.100 & 0.200\\
     19 & 2D & 0.5385 & 0.416 & 0.069 & 0.023 & 0.042 & $ 2.01\times10^{-4} $ & $ 4.02\times10^{-4} $ & 0.125 & 0.250\\
     20 & 2D & 0.4815 & 0.422 & 0.069 & 0.023 & 0.041 & $ 1.99\times10^{-4} $ & $ 3.98\times10^{-4} $ & 0.125 & 0.250\\
     21 & 2D & 0.4286 & 0.427 & 0.069 & 0.023 & 0.040 & $ 1.97\times10^{-4} $ & $ 3.94\times10^{-4} $ & 0.125 & 0.250\\
     22 & 2D & 0.3333 & 0.437 & 0.068 & 0.023 & 0.039 & $ 1.94\times10^{-4} $ & $ 3.88\times10^{-4} $ & 0.125 & 0.250\\
     23 & 2D & 0.2500 & 0.446 & 0.068 & 0.023 & 0.038 & $ 1.91\times10^{-4} $ & $ 3.82\times10^{-4} $ & 0.125 & 0.250\\
     24 & 2D & 0.1765 & 0.455 & 0.067 & 0.022 & 0.037 & $ 1.88\times10^{-4} $ & $ 3.76\times10^{-4} $ & 0.125 & 0.250\\
     25 & 2D & 0.1111 & 0.464 & 0.067 & 0.022 & 0.036 & $ 1.85\times10^{-4} $ & $ 3.70\times10^{-4} $ & 0.125 & 0.250\\
     26 & 2D & 0.0526 & 0.473 & 0.066 & 0.022 & 0.035 & $ 1.83\times10^{-4} $ & $ 3.66\times10^{-4} $ & 0.125 & 0.250\\
     27 & 2D & 0.0256 & 0.477 & 0.066 & 0.022 & 0.034 & $ 1.82\times10^{-4} $ & $ 3.63\times10^{-4} $ & 0.125 & 0.250\\[6pt]
\hline

\end{tabular}
\end{center}

\end{table*}
The SU\,UMa systems Z\,Cha and OY\,Car in outburst have brightness temperatures, $T_{\rm BR}$, in the outer disc  of $\sim$ 6000 -- 7000\,K \nocite{HC85,BBB96}({Horne} \& {Cook} 1985; {Bruch}, {Beele} \& {Baptista} 1996). For the eight SU\,UMa eclipsing systems in Table~\ref{tb:eclipse} we obtain $0.031\,a\,\Omega_{\rm orb}\lesssim c_{\rm s}\lesssim 0.036\,a\,\Omega_{\rm orb}$ at the disc midplane at $R_{3:1}$, assuming a mass transfer rate of  $\dot{M}=10^{-9}\,{\rm M_{\odot}\,yr^{-1}}$ and a fully ionised cosmic mixture of gases. Hence our simulations have a sound speed within a factor of two of that prevailing in reality, a much better match than previously possible.
\begin{table*}
\caption{Parameters of the eight eclipsing SU\,UMa systems: mass ratio, primary mass, total system mass ($M_{1}+M_{2}$), observed disc precession rate (from the measured $\epsilon$ value), dynamical precession rate as calculated from Equation~\ref{eq:dyn} at the location of the 3:1 resonance, inferred pressure contribution to precession ($\omega_{\rm obs}- \omega_{\rm dyn}$) and pitch angle of the spiral wave (see Section~4 for details). Errors on $\omega_{\rm obs}$ and $i_{\rm p}$ correspond to the range of superhump periods observed. In the last but one column the sound speed, in SPH units, is calculated for the midplane of each disc at the 3:1 resonance radius, assuming a mass transfer rate of $10^{-9}\,{\rm M_{\odot}\,yr^{-1}}$ and a fully ionised `cosmic' mixture. References for $M_{1}$ and $M_{\rm t}$ are provided in the final column. Details and references for all other observational data can be found in Table~\ref{tb:epsilon}.}

\label{tb:eclipse}
\begin{center}
\begin{minipage}{\textwidth}
\setcounter{mpfootnote}{0}
\renewcommand{\thempfootnote}{\arabic{mpfootnote}}
\begin{tabular}{llllcccccc}

\hline
System & $q$ & $M_{1}$ & $M_{\rm t}$ & $\omega_{\rm obs}$ & $\omega_{\rm dyn}(R_{3:1})$ & $\omega_{\rm pr}$ & $i_{\rm p}$ & $c_{\rm s}(R_{3:1})$ & Ref\\
& & ($M_{\odot}$) & ($M_{\odot}$) & (${\rm rad\,d^{-1}}$) & (${\rm rad\,d^{-1}}$) & (${\rm rad\,d^{-1}}$) & ($^{\circ}$) & $a\,\Omega_{\rm orb}$ & \\
\hline
\\
OY Car & $0.102 \pm 0.003$ & $0.685 \pm 0.011$ & $0.755 \pm 0.011$ & $2.189^{+0.447}_{-0.181} $ & $3.649$ & $-1.460$ & $12.69^{+2.44}_{-0.70}$ & $0.0334$ & \footnote{\nocite{WHB89}{Wood} {et~al.} (1989)} \\[6pt]
XZ Eri & $0.1098 \pm 0.0017$ & $0.767 \pm 0.018$ & $0.851 \pm 0.018$ & $2.697^{+0.328}_{-0.0035}$ & $4.015$ & $-1.319$ & $13.02^{+1.91}_{-0.16}$ & $0.0321$ & \footnote{\nocite{FDM04b}{Feline} {et~al.} (2004c)}\saveFN\fdmbii\\[6pt]
IY UMa & $0.125 \pm 0.008$ & $0.79 \pm 0.04$ & $0.89 \pm 0.04$ & $2.208^{+0.329}_{-0.120}$ & $3.714$ & $-1.506$ & $11.17^{+1.42}_{-0.41}$ & $0.0322$ & \footnote{\nocite{SPR03}{Steeghs} {et~al.} (2003)}\\[6pt]
Z Cha & $0.1495 \pm 0.0035$ & $0.84 \pm 0.09$ & $0.965 \pm 0.091$ & $3.161^{+0.634}_{-0}$ & $4.282$ & $-1.121$ & $14.31^{+6.85}_{-0}$ & $0.0360$ & \footnote{\nocite{WaH88}{Wade} \& {Horne} (1988)}\\[6pt]
HT Cas & $0.15 \pm 0.03$ & $0.61 \pm 0.04$ & $0.70 \pm 0.04$ & $2.725^{+0}_{-0}$ & $4.343$ & $-1.618$ & $11.62^{+0}_{-0}$ & $0.0347$ & \footnote{\nocite{HWS91}{Horne}, {Wood} \& {Stiening} (1991)}\\[6pt]
DV UMa & $0.1506 \pm 0.0009$ & $1.041 \pm 0.024$ & $1.198 \pm 0.024$ & $2.349^{+0.080}_{-0.286}$ & $3.738$ & $-1.389$ & $10.54^{+0.31}_{-0.92}$ & $0.0314$ & \useFN\fdmbii\\[6pt]
OU Vir & $0.175 \pm 0.025$ & $0.90 \pm 0.19$ & $1.06 \pm 0.19$ & $2.732^{+0.033}_{-0}$ & $4.987$ & $-2.255$ & $8.76^{+0.06}_{-0}$ & $0.0305$ & \footnote{\nocite{FDM04,FDM04err}{Feline} {et~al.} (2004a,b)}\\[6pt]
V2051 Oph & $0.19 \pm 0.03$ & $0.78 \pm 0.06$ & $0.93 \pm 0.07$ & $2.748^{+0.815}_{-0.319}$ & $6.201$ & $-3.453$ & $7.83^{+1.11}_{-0.33}$ & $0.0312$ & \footnote{\nocite{BCH98}{Baptista} {et~al.} (1998)}\\[6pt]

\hline

\end{tabular}
\end{minipage}
\end{center}
\end{table*}

We have also improved on the viscosity used in previous calculations. Our code includes an artificial viscosity term which generates a shear viscosity in the disc,
\begin{equation} \label{eq:sphv}
\nu(r)=\kappa\,\zeta\,c_{\rm s}\,H,
\end{equation}
where $\zeta$ is the dimensionless artificial viscosity parameter. $\zeta=1$ here, and $H$ is the disc scaleheight. $\kappa$ may be found analytically, and for a standard cubic spline kernel in three dimensions, $\kappa=1/10$, whilst in two dimensions, $\kappa=1/8$. The bulk viscosity is fixed to be twice the shear viscosity. Using the Shakura--Sunyaev viscosity parametrisation ($\nu=\alpha\,c_{\rm s}\,H$) \nocite{SS73}({Shakura} \& {Sunyaev} 1973), our simulated 3D and 2D discs have $\alpha(R_{3:1})=0.1$ and $0.125$ respectively. This matches estimates of $\alpha \sim 0.1-0.2$ derived from observation of systems in the high viscosity state \nocite{Sm99}({Smak} 1999).

\subsection{Simulation Results}

\begin{figure*}
\centering
\includegraphics[width=\textwidth]{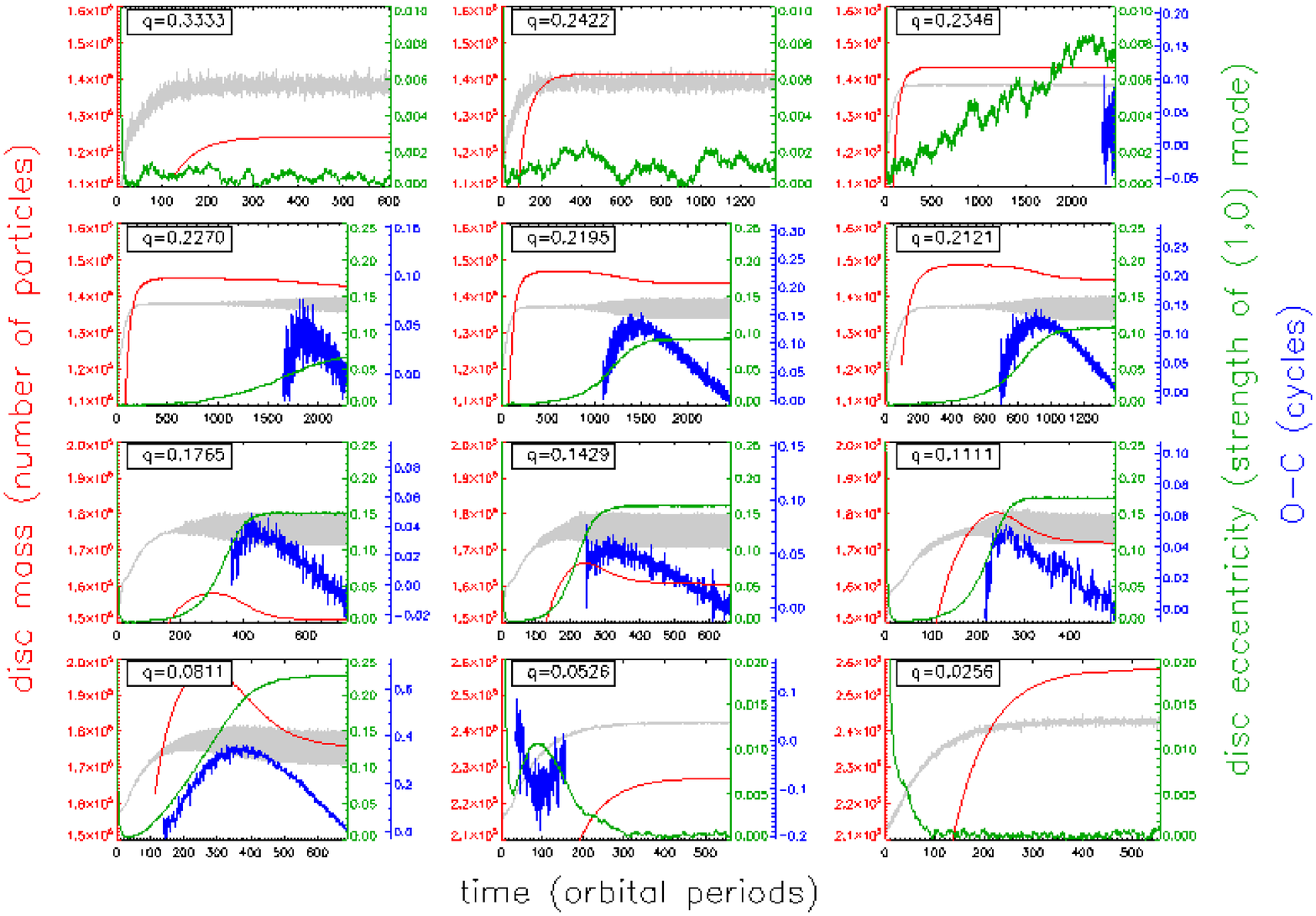}
\caption{Disc mass (red) and disc eccentricity (green) as a function of time for the 3D simulations 1 to 12. Every 5000th and 2000th timestep is plotted respectively. Also plotted is the superhump period (blue) in the form of O-C. This is calculated relative to the mean superhump period from each run, derived from the point at which the superhump signal becomes well-formed. The dissipation (smoothed) from the disc for $r>0.3\,a$ is shown in grey. The mass ratio is shown in the upper left of each panel.}
\label{mass}
\end{figure*}

Table~\ref{tb:sph} summarises our simulations. In some cases these have run for $> 2000$ orbits without reaching mass-transfer equilibrium; one has run for over a year of elapsed time. The rate of energy dissipation (i.e. viscously-generated luminosity with units $M\,a^2\,\Omega_{\rm orb}^3$) from different regions in the disc is recorded each timestep and used to produce simulation lightcurves. $P_{\rm sh}$ was determined from timings of maxima in dissipation in a smoothed lightcurve for the disc region $r>0.3\,a$. Column 8 of Table~\ref{tb:sph} gives the mean value of $\epsilon$ so obtained, over the time in which the system has reached equilibrium where applicable. Figure~\ref{mass} shows the disc evolution for the 3D simulations 1 to 12.
Here the disc mass is taken to be the number of SPH particles in the disc. The disc eccentricity is estimated from the eccentric mode strength, $S_{(1,0)}$, where the strength of the $(k\,\theta-l\,\Omega_{\rm orb}\,t)$ mode, $S_{(k,l)}$, is obtained by Fourier decomposing the simulated disc density distributions in azimuth and time \nocite{L91b,Mu96}({Lubow} 1991b; {Murray} 1996).

\subsubsection{The growth of the eccentricity}
Figure~\ref{mass} shows that initially, the eccentricity grows exponentially.
\begin{figure}
\centering
\includegraphics[width=\columnwidth]{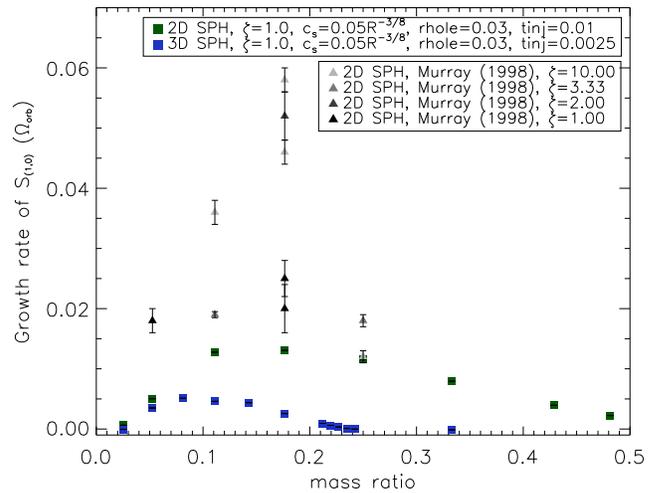}
\caption{Eccentricity growth rate (growth rate of the strength of the $(1,0)$ eccentric mode) as a function of mass ratio for simulations presented in this work, and in previous works (Murray 1998). Simulation parameters are as indicated in the legend.}
\label{growth}
\end{figure}
\nocite{L91}{Lubow} (1991a) found an exponential eccentricity growth rate which is proportional to the square of the mass ratio; in contrast we see a very low growth rate for high mass ratios ($q\gtrsim0.2$) (Figure~\ref{growth}). Our results can be profitably compared with those of \nocite{GO06}{Goodchild} \& {Ogilvie} (2006) who formulated a single equation to describe the resonant excitation, propagation and viscous damping of the eccentricity in a 2D accretion disc. They showed that the resonance may have the effect of locally suppressing the eccentricity which consequently leads to extremely low eccentricity growth rates. We find for $q=0.2195$, for example, a steady state is not reached until $\sim 1700$ orbital periods, and for $q=0.2346$ a steady state is still not achieved after $\sim 2500$ orbital periods.

\subsubsection{The enhanced viscous torques of an eccentric disc}
For all calculations in which the disc (eventually) becomes sufficiently eccentric we see similar behaviour but on different timescales. As Figure~\ref{mass} shows, initially the disc mass builds, exponentially approaching a steady-state value. Then as eccentricity increases, the disc mass tends to a new lower steady state value. This reveals the non-eccentric accretion disc approaching an equilibrium between the tidal removal of angular momentum and the angular momentum added by material from the L1 point. As the disc becomes eccentric it readjusts to a new equilibrium in which tidal removal of angular momentum is more efficient. This is exactly the premise of the TTI model.

Table~\ref{tb:viscous} quantitatively examines this. As a measure of the increase in efficiency of tidal removal of angular momentum in the eccentric disc, we took the decrease in disc mass between maximum and the value it finally reached in equilibrium (column 4 of Table~\ref{tb:viscous}). For a steady state accretion disc the total mass is
\begin{equation}\label{eq:steady}
M_{\rm tot}=\int_{R_*}^{R_{\rm out}} \frac{2 R \dot{M}}{3 \nu(R)}\left[1-\left( \frac{R_*}{R_{\rm out}}\right)^\frac{1}{2}\right]\mathrm{d}R
\end{equation}
\nocite{FKR}({Frank}, {King} \& {Raine} 1985).
The mass transfer rate through ${\rm L_{1}}$ remains constant, and we are comparing the equilibria with and without an eccentric precessing disc. In both cases, the mass accretion at all disc radii must equal the mass transfer rate at ${\rm L_{1}}$. Thus, the change in disc mass can then be related to a change in viscosity by
\begin{equation}
M_{\rm b}-M_{\rm a}=\Delta M={\rm A}\left( \int \frac{f\left( R\right) \mathrm{d}R}{\nu_{\rm b}(R)}-\int \frac{f\left( R\right) \mathrm{d}R}{\nu_{\rm a}(R)}\right),
\end{equation}
where A is a constant, and the subscripts b and a respectively refer to before and after the disc became fully eccentric. Defining
\begin{equation}
\frac{1}{\bar{\nu_b}}=\frac{\int \frac{f\left( R\right) \mathrm{d}R}{\nu_{\rm b}(R)}}{\int f\left( R\right) \mathrm{d}R} ,
\end{equation}
where $\bar{\nu_b}$ is some unknown weighted mean value of $\nu_{\rm b}(R)$, and similarly for $\frac{1}{\bar{\nu_a}}$, then we have
\begin{equation}\label{eq:mass}
\frac{\Delta M}{{\rm A} \int f\left( R\right) \mathrm{d}R}=\left[ \frac{1}{\bar{\nu_b}}-\frac{1}{\bar{\nu_a}}\right].
\end{equation}
This quantity is given in column 5 of Table~\ref{tb:viscous}. The radial dependence of $H$ is $c_{\rm s}(r/\mu)^{1/2}r$, where $\mu=1/(q+1)$. Together with the sound speed prescription that we use, then this allows us, via Equation~\ref{eq:sphv}, to give explicitly the fractional change in viscous torque necessary to bring about the decrease in disc mass we see (column 6 of Table~\ref{tb:viscous}). We assume $R_{\rm out} = R_{\rm tides}$. Here, we use the formulation $R_{\rm tides}\simeq0.9R_1$ \nocite{FKR}({Frank}, {King} \& {Raine} 1985), where $R_1$ is the effective Roche lobe radius \nocite{E83}({Eggleton} 1983), though we note the limitation of this approximation \nocite{MWW00,T07}(e.g. {Murray}, {Warner} \&  {Wickramasinghe} 2000; {Truss} 2007). For runs 6, 7, 9 and 10 we measured the average outer disc radius, averaging the position of the outermost particle as a function of azimuth over a superhump period. We found in each case the difference between this value and $R_{\rm tides}$ to be $<0.006\,a$.
\begin{table}
\caption{Enhanced viscous torque resulting from the accretion disc encounter with the eccentric resonance. The fourth column records the fractional decrease in disc mass before and after the disc becomes fully eccentric. The fifth column gives a measure of the change in viscous torque indicated by this mass decrease as described in the text and given by Equation~\ref{eq:mass}. The final column assumes the radial dependence of viscosity is given by Equation~\ref{eq:sphv}, and records, in that case, the fractional change in viscous torque necessary to bring about the decrease in disc mass.}
\label{tb:viscous}
\begin{center}
\begin{tabular}{llllllllllllll}

Run & 2D/ & $q$ & $\Delta M/M$ & $\frac{1}{\bar{\nu_{\rm b}}}-\frac{1}{\bar{\nu_{\rm a}}}$ & $\nu_{\rm a}/\nu_{\rm b}$\\
  & 3D & & & $1/(a^{2}\Omega_{\rm orb})$ & \\
\hline
\\
      5 & 3D & 0.2195 & 0.0223 & 175 & 1.019 \\
      6 & 3D & 0.2121 & 0.0270 & 211 & 1.023 \\
      7 & 3D & 0.1765 & 0.0464 & 360 & 1.040 \\
      8 & 3D & 0.1429 & 0.0360 & 273 & 1.031 \\
      9 & 3D & 0.1111 & 0.0465 & 353 & 1.040 \\
     10 & 3D & 0.0811 & 0.1063 & 800 & 1.098 \\
     20 & 2D & 0.4815 & 0.0735 & 423 & 1.060 \\
     21 & 2D & 0.4286 & 0.1145 & 665 & 1.097 \\
     22 & 2D & 0.3333 & 0.205 & 1200 & 1.190 \\
     23 & 2D & 0.2500 & 0.267 & 1538 & 1.258 \\
     24 & 2D & 0.1765 & 0.257 & 1392 & 1.231 \\
     25 & 2D & 0.1111 & 0.165 & 823 & 1.127 \\
[6pt]

\hline
\end{tabular}
\end{center}
\end{table}

\subsubsection{The superhump}
\begin{figure}
\centering
\includegraphics[width=\columnwidth]{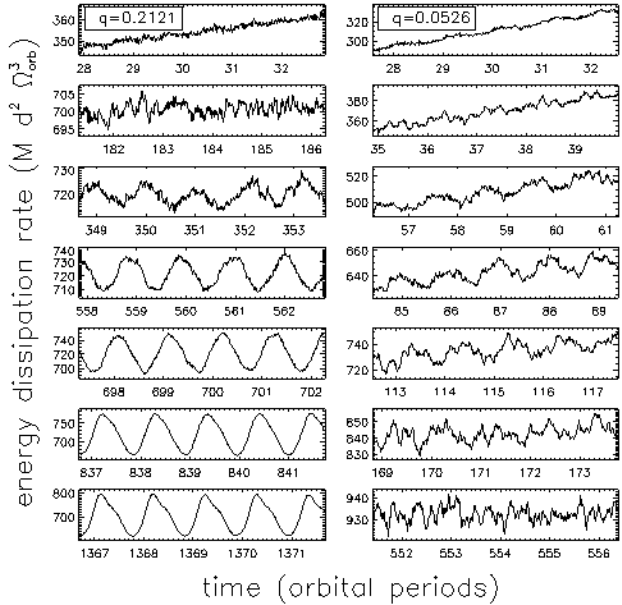}
\caption{Evolution of the simulated lightcurve for runs 6 and 11. Each panel covers 5 orbits. The lightcurve, which has been smoothed somewhat, is given by the rate of energy dissipation for radii $r>0.3\,a$. The mass ratio is shown in the uppermost panel of each.}
\label{detail}
\end{figure}

The development of the superhump for two selected simulations is shown in Figure~\ref{detail}. 
For $q=0.0526$, superhumps were only temporarily present at a time when the eccentricity was highest. For $q=0.2121$ we see that the superhump profile evolves with time as the disc is reaching eccentric equilibrium. In the case of other more extreme mass ratios (smaller values of $q$) there are higher frequency periodic components present in the early development of the superhumps (Smith, PhD thesis in prep.). Our discs accumulate from zero mass with no switch between viscosity states, so the development of their superhump will differ from that of discs observed in superoutburst. 

Variations in the superhump period are displayed in the form of O-C (`observed' minus `calculated') in Figure~\ref{mass}. Intervals of non-zero 2nd derivative in the value of O-C indicate period changes. Generally $P_{\rm sh}$ decreases as the system approaches eccentric steady-state. For $q=0.0526$ a steady-state eccentric disc is not achieved, and the period evolution differs. 

\begin{figure}
\centering
\includegraphics[width=\columnwidth]{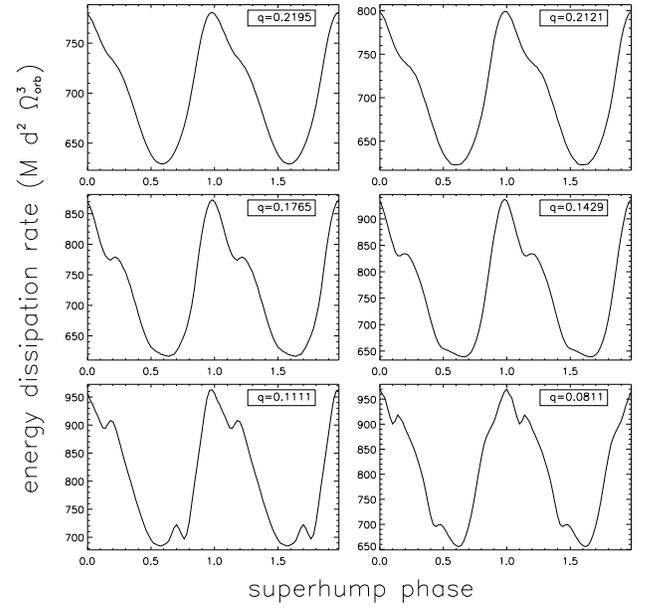}
\caption{Equilibrium lightcurves (energy dissipation rate for radii $r>0.3\,a$) folded on the derived mean superhump period for runs 5 to 10. The superhump cycle is repeated for clarity.}
\label{superhumpfig}
\end{figure}

Each simulated lightcurve was folded on the mean superhump period once equilibrium was reached (Figure~\ref{superhumpfig}). 
These are asymmetric with, for most, a steep rise and slower decline. A secondary hump structure is seen, the profile different for different mass ratios. In general, these superhump profiles resemble those observed in CVs cf. figure~8 of \nocite{PJK95}{Patterson} {et~al.} (1995), figure~4 of \nocite{IKK06}{Imada} {et~al.} (2006) and figure~6 of \nocite{MHN06}{Maehara}, {Hachisu} \& {Nakajima} (2006).

\begin{figure}
\centering
\includegraphics[width=\columnwidth]{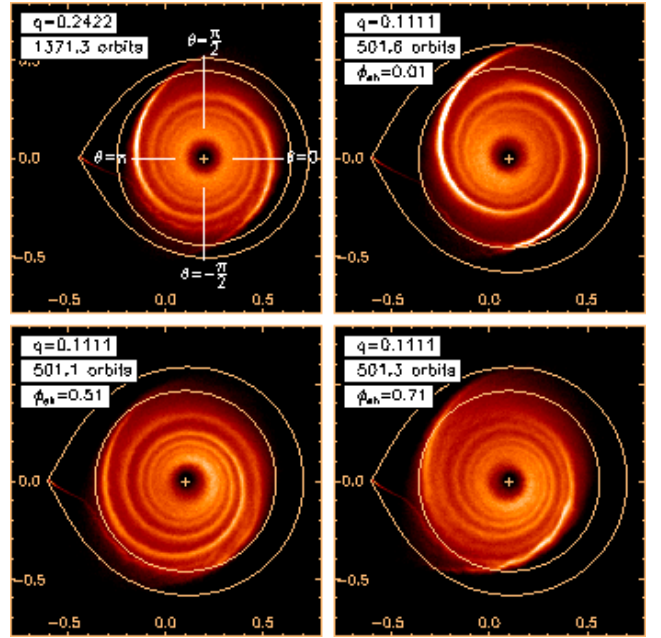}
\caption{Density profiles comparing the non-superhumping disc of run 2 (upper left panel) with the superhumping disc of run 9 at 3 selected superhump phases, $\phi_{\rm sh}$. The coordinates are centred on the binary system centre of mass. The solid line is the primary Roche lobe and the dashed line is the 3:1 resonance radius. The upper left panel also shows our definition of azimuth used in Section~\ref{traj}. The secondary star is at an azimuth of $\pi$ radians with respect to the primary, the position of which is marked with a cross.}
\label{discplots}
\end{figure}

In Figure~\ref{discplots} we show the density distributions of the accretion disc for run 9 at selected superhump phases, and compare these with the density profile of a disc which does not show superhumps. 
The disc is not maximally distorted at the time of maximum energy production ($\phi_{\rm sh}=0$) as might be expected, but at $\phi_{\rm sh}=0.71$. For $q=0.1111$, this coincides with a small secondary peak seen in Figure~\ref{superhumpfig}. The superhumping disc is most similar to the non-superhumping disc at superhump maximum, the most visible difference being in the strongly enhanced spiral density waves in the superhumping disc. Furthermore, the appearance of the spiral density wave changes dramatically over $P_{\rm sh}$. This illustrates the crucial importance of the spiral density waves to the superhump phenomenon, as suggested by \nocite{L91}{Lubow} (1991a) and \nocite{O03}{Osaki} (2003) who considered analytic theory and observations respectively. Comparing superhump maximum with superhump minimum, then we see more a more open spiral structure at superhump maximum with outer reaches showing enhanced density. 
After superhump maximum, differential motion in the superhumping disc causes the spiral to become eccentric and more tightly wrapped with lower density contrast (see the bottom two panels in Figure~\ref{discplots}). Each radius in the disc has its own characteristic Keplerian angular velocity, and eccentric mode precession rate. Differential precession in the eccentric fluid disc cannot persist, because this would cause widespread orbit crossings. This converging fluid motion causes the strong spiral shock shown in the upper right panel of Figure~\ref{discplots}, and it is the dissipation associated with this shock which powers the observed superhump. Figure~\ref{discplots} illustrates the mechanism by which spiral waves power superhumps.

\subsubsection{2D versus 3D}

\begin{figure*}
\centering
\includegraphics[width=\textwidth]{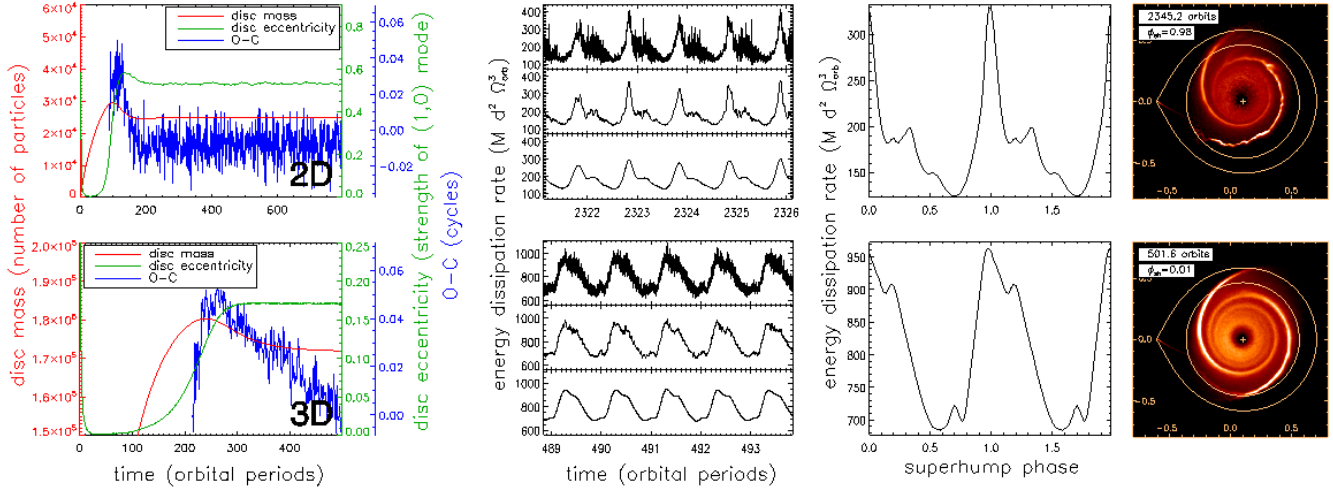}
\caption{Comparison of results for 2D and 3D simulations with $q=0.1111$, runs 24 and 9 respectively. The following are shown above and below respectively for each: the disc evolution, details of the simulated lightcurves, folded superhump lightcurves and density profiles. Here symbols are as in Figure~\ref{discplots}, and the colour scale used is the same for both the 2D and 3D discs. The simulated lightcurves cover 5 orbital cycles once the disc mass and eccentricity had stabilised. The raw lightcurve is shown in the top panel and increasingly smoothed lightcurves below.}
\label{compare}
\end{figure*}

In Figure~\ref{compare} we compare the results of $q=0.1111$ calculations in 2D and 3D. 
In the 2D simulations the accretion disc achieves a far higher eccentricity more quickly. This in turn affects the superhump profile. We see from Table~\ref{tb:sph} that these 2D accretion discs (runs 19 to 27) can become eccentric at much larger values of mass ratio than in the 3D simulations, and well beyond the bounds that theory suggests, indicating that three dimensions and the high mass resolution ($\Delta\,t=0.0025$) is necessary to accurately simulate the processes involved. In this case, (3D, $\Delta\,t=0.0025$) no disc develops superhumps where $q\gtrsim0.24$. We also see that for the same value of $q$, the 2D discs precess at smaller rates than the 3D discs; compare for example run 24 with run 7.

\section{Eccentricity distribution in accretion discs} \label{ecc}

An eccentric accretion disc underlies the superhump phenomenon, but how eccentric does the disc become? How does the eccentricity vary throughout the disc, and how does it change with superhump phase?

We plotted an estimate of the disc eccentricity in our simulations in Figure~\ref{mass}. Calculating the strength of the (1,0) mode takes into consideration the disc as a whole. For those 3D discs which reached equilibrium and showed superhumps, final values of eccentricity $e$ between $\sim0.09$ and $\sim0.23$ were seen, with more extreme mass ratios harbouring more eccentric discs. This agrees with modelling of line profiles in AM\,CVn assuming a constant eccentricity throughout the disc: \nocite{PHS93}{Patterson}, {Halpern} \&  {Shambrook} (1993c) found $e=0.1-0.2$. In their analytic study \nocite{GO06}{Goodchild} \& {Ogilvie} (2006) examined the spatial distribution of eccentricity, but explicitly avoided examination of the behaviour on the orbital timescale. Their eccentricity distribution was locally suppressed by the presence of dynamical resonances. Next we examine the eccentricity distributions of our simulated discs, using two complementary methods to characterise their spatial and temporal variation.

\subsection{Eccentricity distribution from particle trajectories}
\label{traj}

An instantaneous snapshot of the radial eccentricity distribution was found by projecting the elliptical orbit of each particle in the disc using its position and velocity. Particles in the mass transfer stream were discounted. Each particle was assigned a radius given by an average along this elliptical path, weighted by the time spent at each radius. This method calculates the trajectories particles would have if they orbited an isolated primary star. This simplification may introduce certain artefacts into the calculated eccentricity distributions, but does give an adequate approximation.

\begin{figure*}
\centering
\includegraphics[width=\textwidth]{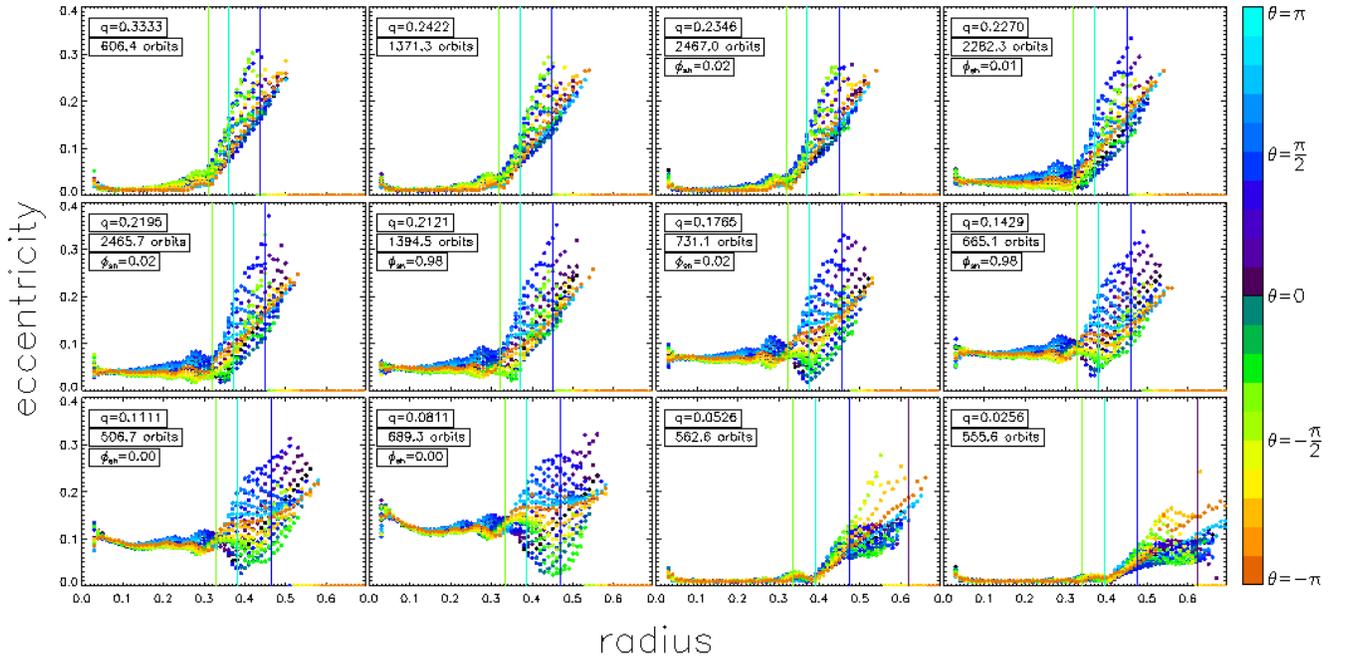}
\caption{Instantaneous eccentricity distribution plotted as a function of radius for 3D runs 1 to 12, once equilibrium is reached (see text for details). Each point represents the average of all disc particles in each of 20 equal azimuthal sections and binned in radius. These are colour coded by their azimuthal position in the disc, as defined in Figure~\ref{discplots}, according to the scale on the right. The time is shown in the upper right of each panel, together with the superhump phase where applicable.}
\label{eccvrad}
\end{figure*}

Figure~\ref{eccvrad} shows the results for each of the 3D simulations 1 to 12, at superhump maximum where applicable. 
In this and subsequent similar figures, each point represents the mean eccentricity of particles, obtained in the manner described above, in each of a number of azimuthal and radial bins of size $0.1\pi$\,rad and $0.1\,a$ respectively. Overplotted on these figures are the 3:1, 4:1 and 5:1 resonance radii from right to left respectively. For simulations 11 and 12 the 2:1 resonance is also shown. Each particle is colour coded according to the azimuthal position of the particles it represents, as shown in the key, where $\theta$ is the angle defined in Figure~\ref{discplots}. Purple and blue particles are approaching the mass donor star, while green and yellow particles are receding from it. The eccentricity distributions at different azimuths are clearly distinct. Caution is required, however, in interpreting the radii assigned in this section. The particles at $r > R_{3:1}$ are concentrated at azimuths close to the x-axis, where in fact the disc edge is relatively close to the compact object: these particles have larger radii assigned to them than those found further out at other azimuths because they have relatively circular projected orbits.

For all the discs with $ q \ge 0.0811$ the eccentricity in the outer disc has a wide spread around $ e \sim 0.2$. This agrees reasonably with the eccentricities found in the outer discs of OY\,Car: \nocite{HMB92}{Hessman} {et~al.} (1992) found $e=0.38\pm0.10$; and IY\, UMa: \nocite{PKJ00}{Patterson} {et~al.} (2000a) estimate $e=0.29\pm0.06$. For $ 0.2121 \leq q \leq 0.3333 $ the eccentricity distributions are low for the disc within the 5:1 resonance radius, then increase at larger radii to $e$ up to $\sim 0.3$, this apparent eccentricity likely due to tidal distortions in the outer disc. For this range of $q$, if every particle were individually plotted, the eccentricity distribution outside $R_{5:1}$ is ``double horned", with a concentration of points near the upper and lower edges of the distribution and a dearth of points midway between the envelope. These two ``horns" correspond to the halves of the disc which are approaching and receding from the mass donor star.

Figure~\ref{eccvrad} demonstrates that the body of the disc within the 5:1 resonance appears eccentric for $0.081 \leq q \leq 0.2195$. For these discs the entire mass appears to partake in the eccentric instability. This roughly flat region of the eccentricity distribution has an eccentricity dependent on $q$, increasing to $\sim 0.12$ as we move to smaller $q$.

\begin{figure}
\centering
\includegraphics[width=\columnwidth]{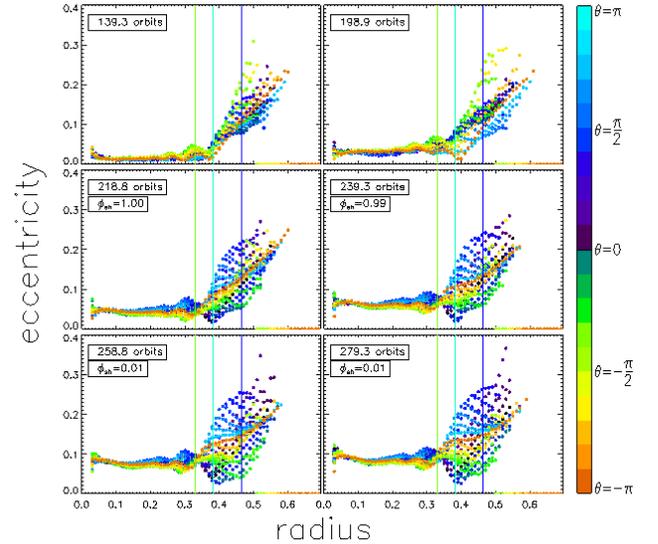}
\caption{Development of accretion disc radial eccentricity distribution for run 9. Azimuthal colour-coding is as in Figure~\ref{eccvrad}.}
\label{eccdevelop}
\end{figure}

Figure~\ref{eccdevelop} shows how the eccentricity distribution develops in run 9 as the disc becomes eccentric. 
Before superhumps set in (top left panel) the eccentricity distribution resembles those seen in the less extreme mass ratio systems i.e. the top row of panels in Figure~\ref{eccvrad}. The remaining 5 panels of Figure~\ref{eccdevelop} span the interval where the disc eccentricity is growing and the disc is emptying as it approaches the new mass transfer equilibrium caused by the enhanced viscous torque (c.f. Figure~\ref{mass}). In each case superhump maximum is plotted. Within $R_{5:1}$ the eccentricity of the body of the disc grows, while the range of eccentricities at any given radius remains more or less unchanged. Outside $R_{5:1}$ the eccentricity distribution becomes more scattered, as the range of eccentricities at any given radius increases. The last panel of Figure~\ref{eccdevelop} shows the distribution when the strength of the (1,0) mode approaches maximum, but mass equilibrium is not yet reached. It resembles the $q=0.1111$ panel in Figure~\ref{eccvrad}, which occurs almost 230 orbits later, after mass equilibrium is established.

\begin{figure}
\centering
\includegraphics[width=\columnwidth]{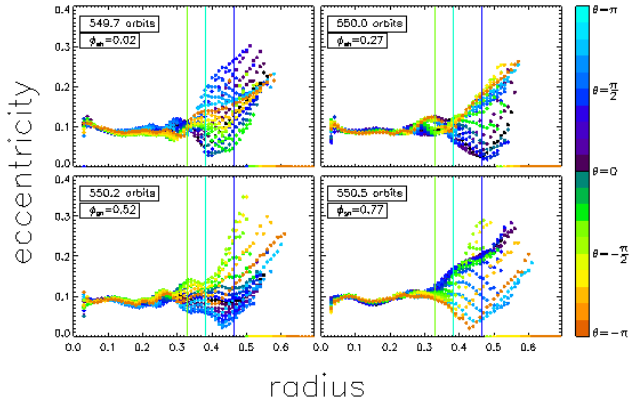}
\caption{As Figure~\ref{eccvrad}, showing the changing radial eccentricity distribution with superhump phase, for run 9.}
\label{eccpsh}
\end{figure}

The eccentricity distribution is shown as a function of superhump phase in Figure~\ref{eccpsh}. 
At some phases and radii there is little spread in $e$, while at other phases there is a large spread in $e$ at the same radius. This behaviour may be related to the cusps at the resonance points discussed by \nocite{GO06}{Goodchild} \& {Ogilvie} (2006), but their treatment explicitly excluded the behaviour of the disc on timescales comparable to or shorter than the orbital period. Numerical simulation facilitates examination of the flexing of the disc during a single orbit, which is important as it is this flexing which gives rise to the modulation in viscously-generated or reprocessed light which constitutes an observed superhump. The flexing of the disc over the superhump period appears most dramatic at radii between $R_{5:1}$ and $R_{3:1}$, in accordance with Pearson's (2006) point that the dynamic precession rate at the 4:1 resonance actually agrees far better with the observations.

Figure~\ref{eccpsh} neatly illustrates the changes which occur as the binary's gravitational potential moves relative to the disc. Even as far in as r=0.15a we see the more eccentric particles belonging to the approaching side of the disc at superhump maximum and the receding part of the disc at superhump minimum.

\subsection{Eccentricity from the mass distribution}

How good a representation of the disc eccentricity are the distributions we calculated in section~\ref{traj}? If we were to follow a single particle as it orbits in the accretion disc it would not move on the elliptical orbit we extrapolated from its instantaneous velocity and position. We now use an alternative way of examining the eccentricity of the accretion disc, looking at the mass distribution.

\begin{figure*}
\centering
\includegraphics[width=\textwidth]{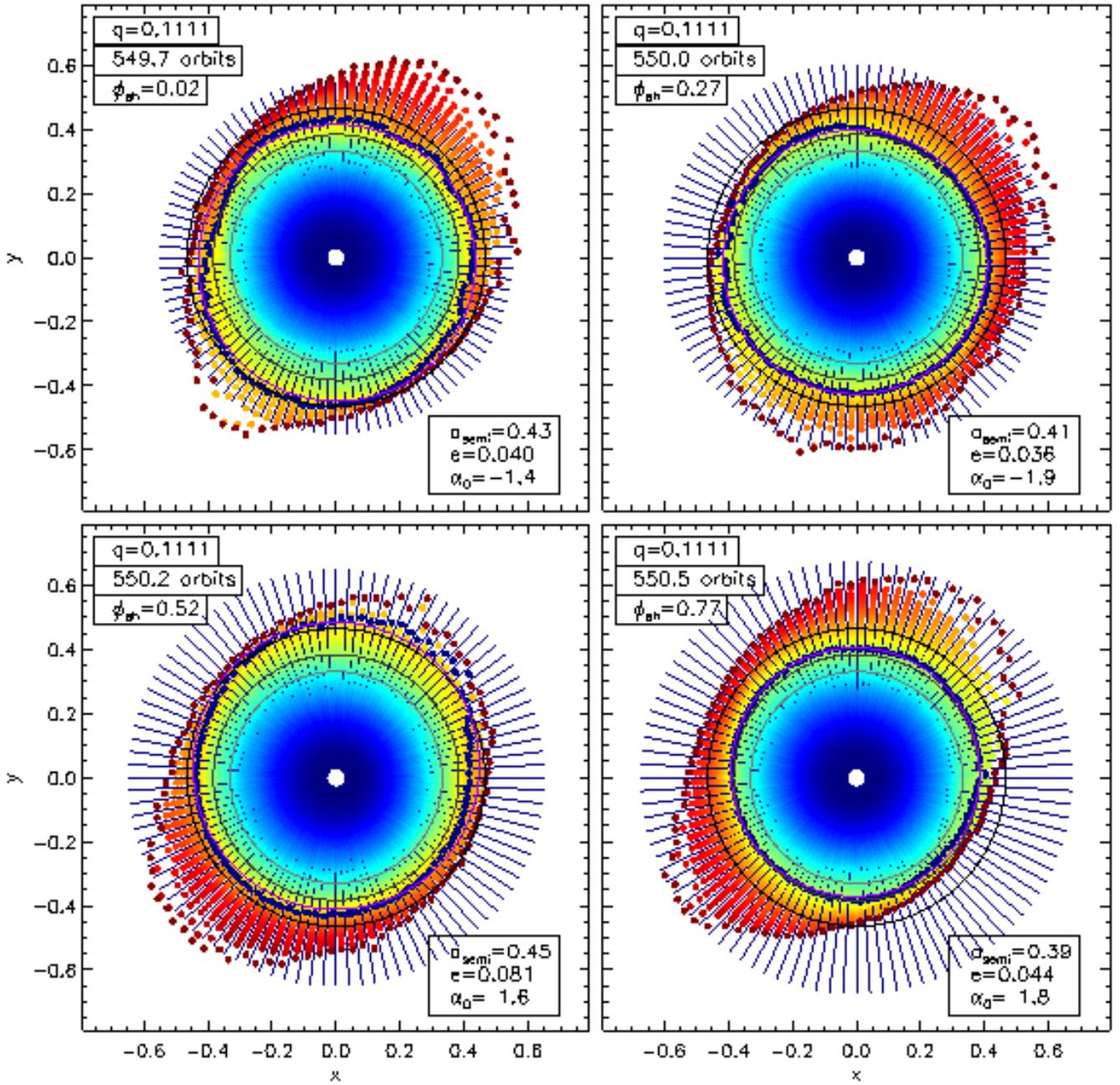}
\caption{Pictorial representation of the mass distribution in the disc for run 9 at four superhump phases. The disc is split into 100 equal azimuthal sections (dashed lines). Each colour graduation represents `contours' of particle numbers. Moving outwards, each contour represents an increase of 24 particles in each azimuthal section (90 contours in all). Further details are described in the text. The outermost dark red points mark the positions of the outermost particles in each azimuthal section. Also plotted are three circles at the 3:1, 4:1 and 5:1 resonance radii moving inward respectively (black, dark and light grey) which also act as a guide to the eye to show the non-circularity of the contours.}
\label{disckl900}
\end{figure*}

The disc in run 9 was split into 100 azimuthal sections. For each, we recorded the number of particles contained within each radial step outwards. In Figure~\ref{disckl900}, we show the contour maps which result. 
Every step of 24 particles was recorded and colour-coded, so each colour represents a `contour' of enclosed mass. To each contour we fitted an ellipse which has one focus at the WD. The fitted parameters are the semi-major axis, $a_{\rm semi}$, the eccentricity, $e$, and the angle the semi-major axis makes with the positive $x$-axis in an anti-clockwise direction, $\alpha_{0}$, measured in radians. This is analogous to our definition of $\theta$ in Section~\ref{traj}. For the outermost complete contour, that is the outermost contour for which each azimuthal section is represented, marked in dark blue, these fitted parameters are displayed in the bottom right of each panel in Figure~\ref{disckl900} and apply to the overplotted magenta ellipse.

Figure~\ref{disckl900} allows us to look at the disc mass distribution in a more quantitative way than simply looking at the density distribution. The eye is drawn to the pronounced non-axisymmetry outside the 3:1 resonance, though this mass constitutes less than 9 per cent of the mass in the disc, and is the lowest density region.  This mass contributes only 4 per cent of the total dissipation, and this contribution to the modulation on $P_{\rm sh}$ is not in phase with the overall superhump. The region of the disc within $R_{3:1}$ is overwhelmingly responsible for generating the dissipation-powered superhump in the simulations.

\begin{figure}
\centering
\includegraphics[width=\columnwidth]{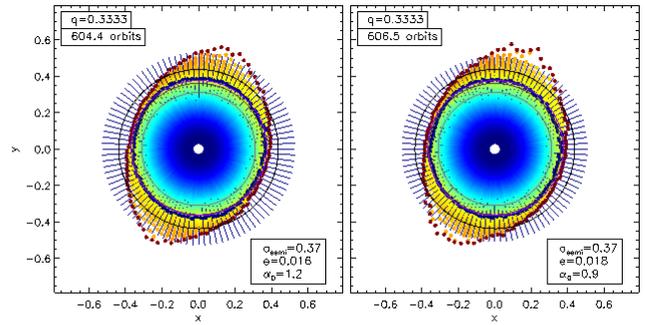}
\caption{As Figure~\ref{disckl900} for the non-superhumping disc of run 1.}
\label{disckl750}
\end{figure}

The four panels in Figure~\ref{disckl900} are equally spaced in superhump phase. It is noticeable that the outermost disc does not simply precess as the binary frame moves. Instead the flexure of the disc combines with the orbital motion of the binary frame to leave the outer edge of the disc almost fixed in the binary frame between $\phi_{sh} = 0.02$ and $\phi_{sh}  = 0.27$; similarly the outer edge of the disc remains almost the same in the two lower panels at $\phi_{sh} =0.52$ and $\phi_{sh} =0.77$. Figure~\ref{disckl750} shows that a non-superhumping disc is extended at similar azimuths. 
These extended disc edges are analogous to the raised tides in Earth's oceans. As Figure~\ref{disckl750} shows, this effect produces an elliptical shape centred on the primary. This illustrates the distinction that needs to be made between different contributions to the eccentricity distributions found here and in Section~\ref{traj}. In the outer disc tidal distortions are important, and are present in discs at all mass ratios. This is to be distinguished from the (1,0) eccentric mode eccentricity found in the superhumping discs which has the primary at one focus. \nocite{T07}{Truss} (2007) finds that for discs which have not yet become tidally unstable, the exact azimuth of the extended `wing' depends on the mass ratio, viscosity parameter and sound speed of the gas (the major axis of the outer streamline moves clockwise with decreasing $q$, or increasing $\alpha$ and $c_{\rm s}$).
The interior regions of the disc (e.g. around $R_{\rm 4:1}$) more closely approximate simple relative motion between a slowly apsidally precessing disc orientation and a rapidly moving orbital frame.

\begin{figure}
\centering
\includegraphics[width=\columnwidth]{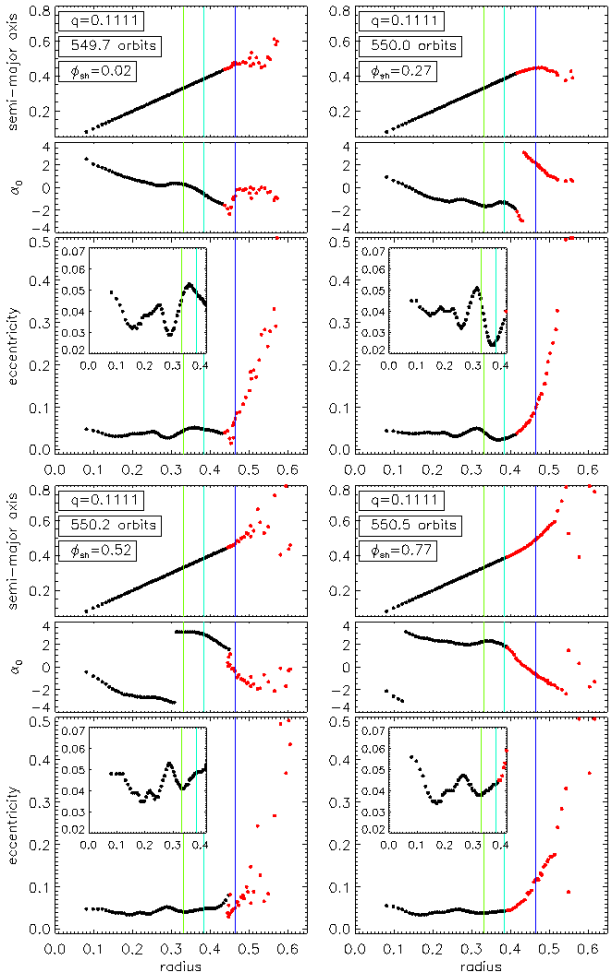}
\caption{Fitted ellipse parameters for each `contour' in Figure~\ref{disckl900}, as a function of radius. See Figure~\ref{disckl900} and text for details. Results for four superhump phases are presented corresponding to the four panels in Figure~\ref{disckl900}. Vertical lines indicate the 3:1, 4:1 and 5:1 resonance radii from right to left respectively. Red points indicate contours which are not complete.}
\label{discfit900}
\end{figure}

Figure~\ref{discfit900} shows the fitted ellipse parameters for the four phases shown in Figure~\ref{disckl900}.
The average radius of all points making up the contour was used. The general trends in Figs.~\ref{discfit900} and \ref{disckl900} are seen in analogous plots for $q=0.1765$, including the peaks and troughs in the radial distribution of the eccentricity parameter. Within $R_{3:1}$, the orientation of the semi-major axis, $\alpha_{0}$, changes systematically over the superhump period and in the opposite sense to the motion of the gas in the disc i.e. precession of the slowly-moving disc as viewed from the rapidly rotating binary reference frame.

Figure~\ref{discfit900} is complementary to Figure~\ref{eccpsh}. Figure~\ref{eccpsh} shows the eccentricity distributions deduced from instantaneous velocities, while Figure~\ref{discfit900} shows the eccentricity distribution deduced from the instantaneous mass distribution. Since the disc is continuously flexing in a complex way, these are not the same. The disc motions can be described as superpositions of the S(k,l) modes, and the resonance radii act as nodes and antinodes in the complex standing wave dynamics the disc executes over a full precession period. To summarise contributions to these disc motions we compare, in table \ref{tb:sph}, the strengths of the $(1,0)$ and $(2,2)$ modes.

\begin{figure}
\centering
\includegraphics[width=\columnwidth]{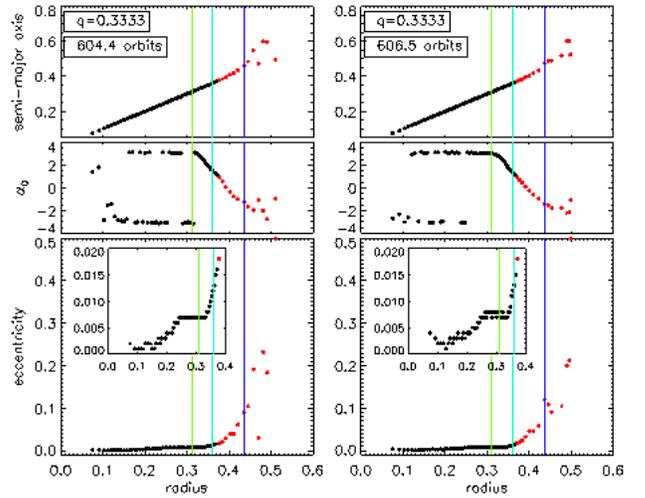}
\caption{As Figure~\ref{discfit900} and corresponding to the panels in Figure~\ref{disckl750}.}
\label{discfit750}
\end{figure}

For comparison, we show the equivalent of Figures~\ref{disckl900} and~\ref{discfit900} for $q=0.3333$ in Figures~\ref{disckl750} and~\ref{discfit750}, a disc which does not show superhumps. 
Here, as we would expect, the mass distribution remains approximately constant over time, and the eccentricity is much lower.

\section{Period Excess versus Mass Ratio: Drawing together observation, theory and simulation} \label{eqcompare}

\begin{figure*}
\centering
\includegraphics[angle=90,width=\textwidth]{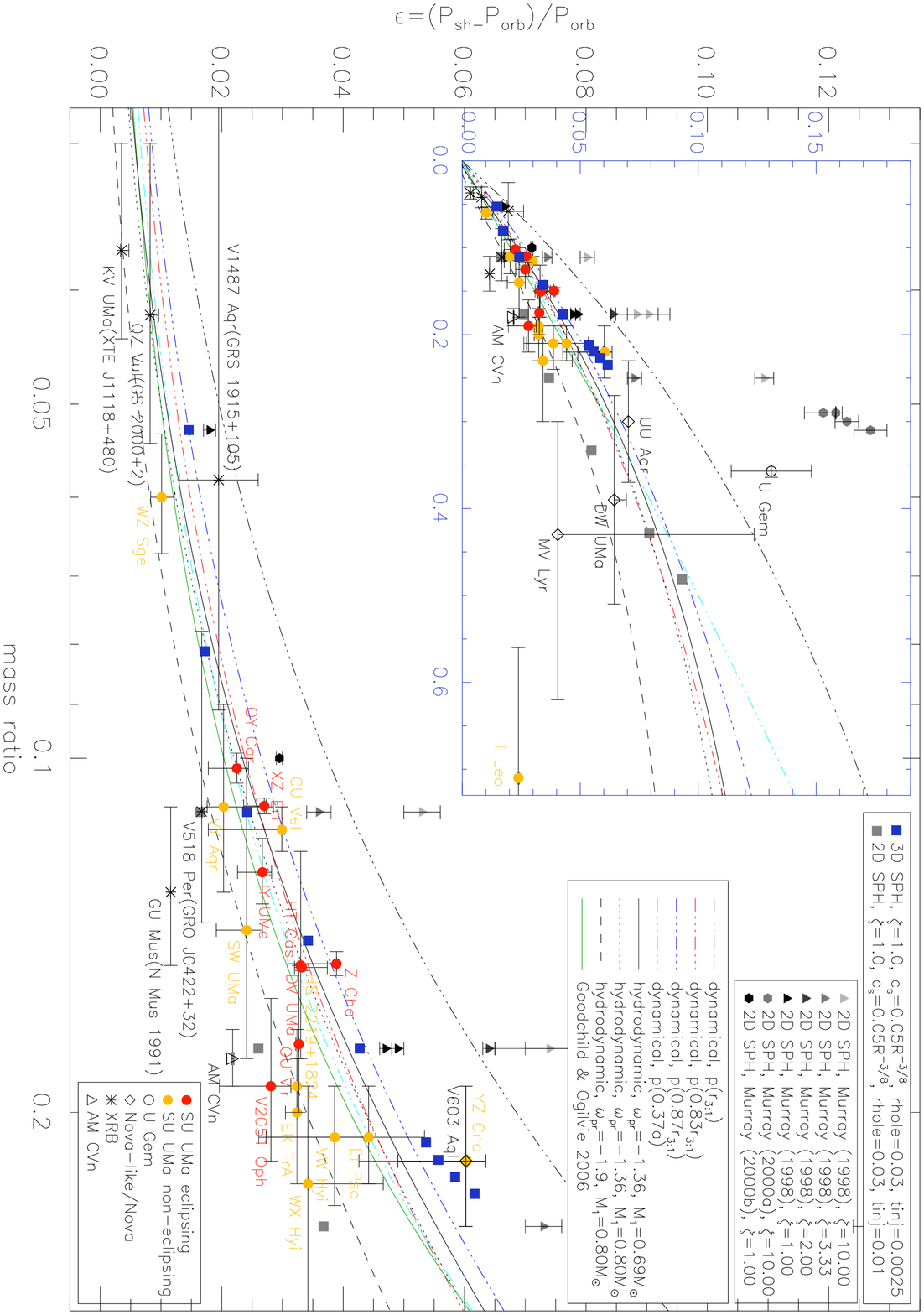}
\caption{Superhump period excess plotted as a function of binary mass ratio for both observed systems and for SPH simulation. Also plotted are the dynamical and hydrodynamical theoretical predictions. Simulations presented in this paper are displayed as filled squares, as detailed in the top right legend. The second legend down refers to previous works by Murray. The third legend refers to theoretical predictions, and the legend for observational data is in the bottom right. The inset shows data over a large range of $q$, whilst the main panel focuses on the data at low $q$ where most of the points are clustered, and is plotted on a logarithmic $x$-axis.}
\label{eq}
\end{figure*}

If a reliable relationship between $\epsilon$ and $q$ can be deduced, this would be immensely useful. $\epsilon$ can be easily measured using relatively modest equipment, while the mass ratio $q$ is more fundamental and less easily determined. \nocite{PKH05}{Patterson} {et~al.} (2005) fitted observations of eclipsing systems with $\epsilon=0.18\,q+0.29\,q^2$.
Figure~\ref{eq} collates observation, theory and simulations of positive superhumps.
All observed systems with $q$ determined by some means independent of $\epsilon$ are plotted. Errors in $q$ are formal errors given by the authors and do not necessarily reflect the uncertainty in the method. The eclipsing systems (red circles) should therefore be given more weight. We note, however, the scatter of the eclipsing systems seems typical of the scatter of the other points. Error bars for $\epsilon$ denote the range of values observed rather than errors in individual values, except in cases where only one measurement has been made. Observational data is tabulated in Table~\ref{tb:epsilon}.

\begin{table*}
\caption{Observed superhump systems with independently determined mass ratio. All periods are given in days. Columns 5 and 6 list the minimum and maximum observed superhump periods respectively. Column 7 gives the superhump excess where the errors indicate the range of values as calculated from min $P_{\rm sh}$ and max $P_{\rm sh}$. In column 8, the object type, (E) denotes a WD eclipsing system, while (e) denotes a system which shows eclipse of the accretion stream/disc impact region only.}
\label{tb:epsilon}

\begin{minipage}{\textwidth}
\begin{scriptsize}
\begin{center}
\setcounter{mpfootnote}{0}
\renewcommand{\thempfootnote}{\arabic{mpfootnote}}
\begin{tabular}{llllllllc}
\hline
System & $q$ & $P_{\rm orb}$ & $P_{\rm sh}$ & min $P_{\rm sh}$ & max $P_{\rm sh}$ & $\epsilon$ & Type & Ref \\
\hline
\\
OY Car & $ 0.102(3) $ & $ 0.0631209180(2) $ & $ 0.06454(2) $ & 0.064245 & 0.06466 & $ 0.0225 _ { - 0.0047 } ^ { + 0.0019 } $ & SU UMa(E) & \footnote{\nocite{WHB89}{Wood} {et~al.} (1989)}$^,$\footnote{\nocite{PHN99}{Pratt} {et~al.} (1999)}$^,$\footnote{\nocite{BBB96}{Bruch} {et~al.} (1996)}$^,$\footnote{\nocite{Sc86}{Schoembs} (1986)}$^,$\footnote{\nocite{PBG93}{Patterson} {et~al.} (1993a)}\saveFN\pbg \\[3pt]
XZ Eri & $ 0.1098(17) $ & $ 0.061159491(5) $ & $ 0.062808(17) $ & 0.062603 & 0.06283 & $ 0.0270 _ { - 0.0034 } ^ { + 0.0003 } $ & SU UMa(E) & \footnote{\nocite{FDM04b}{Feline} {et~al.} (2004c)}\saveFN\fdmbi$^,$\useFN\fdmbi$^,$\footnote{\nocite{UKI04}{Uemura} {et~al.} (2004)}\saveFN\uki$^,$\useFN\uki$^,$\footnote{\nocite{PKH05}{Patterson} {et~al.} (2005)}\saveFN\pkh \\[3pt]
IY UMa & $ 0.125(8) $ & $ 0.07390897(5) $ & $ 0.07588(1) $ & 0.07558 & 0.07599 & $ 0.0267 _ { - 0.0041 } ^ { + 0.0015 } $ & SU UMa(E) & \footnote{\nocite{SPR03}{Steeghs} {et~al.} (2003)}\saveFN\spr$^,$\useFN\spr$^,$\footnote{\nocite{UKM00}{Uemura} {et~al.} (2000)}$^,$\footnote{\nocite{PKJ00}{Patterson} {et~al.} (2000a)}\saveFN\pkj$^,$\useFN\pkj \\[3pt]
Z Cha & $ 0.1495(35) $ & $ 0.074499  $ & $ 0.07740 $ & 0.0768 &  & $ 0.0389 _ { - 0.0080 } ^ { + 0 } $ & SU UMa(E) & \footnote{\nocite{WHB86}{Wood} {et~al.} (1986)}$^,$\footnote{\nocite{BJO02}{Baptista} {et~al.} (2002)}$^,$\footnote{\nocite{WO'D88}{Warner} \& {O'Donoghue} (1988)}\saveFN\wod$ ^,$\useFN\wod \\[3pt]
HT Cas & $ 0.15(3) $ & $ 0.07364720309(7) $ & $ 0.076077  $ &  &  & $ 0.0330           $ & SU UMa(E) & \footnote{\nocite{HWS91}{Horne} {et~al.} (1991)}$^,$\footnote{\nocite{FDM05}{Feline} {et~al.} (2005)}$^,$\footnote{\nocite{ZRN86}{Zhang}, {Robinson} \& {Nather} (1986)} \\[3pt]
DV UMa & $ 0.1506(9) $ & $ 0.0858526521(14) $ & $ 0.08870(8) $ & 0.0886 & 0.08906 & $ 0.0332 _ { - 0.0012 } ^ { + 0.0042 } $ & SU UMa(E) & \useFN\fdmbi$^,$\useFN\fdmbi$^,$\footnote{\nocite{PVS00}{Patterson} {et~al.} (2000b)}\saveFN\pvs$^,$\useFN\pvs$^,$\useFN\pvs \\[3pt]
OU Vir & $ 0.175(25) $ & $ 0.072706113(5) $ & $ 0.07508(9) $ & 0.07505 &  & $ 0.0327 _ { - 0.0005 } ^ { + 0 } $ & SU UMa(E) & \footnote{\nocite{FDM04}{Feline} {et~al.} (2004a)}\saveFN\fdm$^,$\useFN\fdm$^,$\useFN\pkh$^,$\footnote{\nocite{KNM03}{Kato} {et~al.} (2003)}\saveFN\knm \\[3pt]
V2051 Oph & $ 0.19(3) $ & $ 0.0624278634(3) $ & $ 0.06418(16) $ & 0.06365 & 0.06439 & $ 0.0281 _ { - 0.0085 } ^ { + 0.0033 } $ & SU UMa(E) & \footnote{\nocite{BCH98}{Baptista} {et~al.} (1998)}$^,$\footnote{\nocite{BBB03}{Baptista} {et~al.} (2003)}$^,$\footnote{\nocite{PTK03}{Patterson} {et~al.} (2003)}\saveFN\ptk$^,$\useFN\ptk$^,$\useFN\ptk \\[3pt]
WZ Sge & $ 0.060(7) $ & $ 0.0566878460(3) $ & $ 0.05726(1) $ & 0.05716 & 0.05738 & $ 0.0101 _ { - 0.0018 } ^ { + 0.0021 } $ & WZ Sge(e) & \footnote{\nocite{SWL02}{Skidmore} {et~al.} (2002)}$^,$\footnote{\nocite{PRK98}{Patterson} {et~al.} (1998)}$^,$\footnote{\nocite{IUM02}{Ishioka} {et~al.} (2002)}$^,$\useFN\pbg$^,$\footnote{\nocite{PMR02}{Patterson} {et~al.} (2002a)} \\[3pt]
VY Aqr & $ 0.11(2) $ & $ 0.06309(4) $ & $ 0.06437(9) $ & 0.0642 & 0.06489 & $ 0.0203 _ { - 0.0027 } ^ { + 0.0082 } $ & SU UMa & \footnote{\nocite{A94}{Augusteijn} (1994)}\saveFN\abc$^,$\footnote{\nocite{TT97}{Thorstensen} \& {Taylor} (1997)}$^,$\useFN\pbg$^,$\useFN\pbg$^,$\useFN\pbg \\[3pt]
CU Vel & $ 0.115(5) $ & $ 0.0785(2) $ & $ 0.08085(3) $ & 0.0799 &  & $ 0.0299 _ { - 0.0121 } ^ { + 0 } $ & SU UMa & \footnote{\nocite{MD96}{Mennickent} \& {Diaz} (1996)}\saveFN\md$^,$\useFN\md$^,$\useFN\knm$^,$\footnote{\nocite{V81}{Vogt} (1981)} \\[3pt]
SW UMa & $ 0.14(4) $ & $ 0.056815(1) $ & $ 0.058182(7) $ & 0.05790 & 0.05833 & $ 0.0241 _ { - 0.0050 } ^ { + 0.0026 } $ & SU UMa & \footnote{\nocite{Sh83}{Shafter} (1983)}\saveFN\sh$^,$\footnote{\nocite{HS88}{Howell} \& {Szkody} (1988)}$^,$\footnote{\nocite{NBK98}{Nogami} {et~al.} (1998)}$^,$\footnote{\nocite{SOK97}{Semeniuk} {et~al.} (1997)}$^,$\footnote{\nocite{RSH87}{Robinson} {et~al.} (1987)} \\[3pt]
HS 2219+1824 & $ 0.19(1) $ & $ 0.0599  $ & $ 0.06184  $ &  &  & $ 0.0324           $ & SU UMa & \footnote{\nocite{R-GGH05}{Rodr{\'{\i}}guez-Gil} {et~al.} (2005)}\saveFN\rggh$^,$\useFN\rggh$^,$\useFN\rggh \\[3pt]
EK TrA & $ 0.20(3) $ & $ 0.06288(5) $ & $ 0.06492(10) $ & 0.0648 &  & $ 0.0324 _ { - 0.0019 } ^ { + 0 } $ & SU UMa & \footnote{\nocite{MenA98}{Mennickent} \& {Arenas} (1998)}\saveFN\mena$^,$\useFN\mena$^,$\useFN\mena$^,$\footnote{\nocite{VS80}{Vogt} \& {Semeniuk} (1980)} \\[3pt]
EI Psc & $ 0.21(2) $ & $ 0.044572(2) $ & $ 0.04654  $ & 0.04579 &  & $ 0.0442 _ { - 0.0169 } ^ { + 0 } $ & SU UMa & \footnote{\nocite{TFP02}{Thorstensen} {et~al.} (2002)}$^,$\footnote{\nocite{UKI02}{Uemura} {et~al.} (2002b)}$^,$\footnote{\nocite{SKB02}{Skillman} {et~al.} (2002)}\saveFN\skb$^,$\useFN\skb \\[3pt]
VW Hyi & $ 0.21 ^{+0.03}_{-0.02} $ & $ 0.074271038(14) $ & $ 0.07714(5) $ & 0.07621 & 0.07824 & $ 0.0386 _ { - 0.0125 } ^ { + 0.0148 } $ & SU UMa & \footnote{\nocite{SHH06}{Smith}, {Haswell} \& {Hynes} (2006)}\saveFN\shh$^,$\footnote{\nocite{vADG87}{van Amerongen} {et~al.} (1987)}\saveFN\vadg$^,$\useFN\vadg$^,$\footnote{\nocite{V83}{Vogt} (1983)}\saveFN\vbc$^,$\useFN\vbc \\[3pt]
YZ Cnc & $ 0.22  $ & $ 0.0868(2) $ & $ 0.09204  $ & 0.0905 &  & $ 0.0604 _ { - 0.0178 } ^ { + 0 } $ & SU UMa & \footnote{\nocite{SH88}{Shafter} \& {Hessman} (1988)}\saveFN\sha$^,$\useFN\sha$^,$\footnote{\nocite{P79}{Patterson} (1979)}\saveFN\pat$^,$\useFN\pat \\[3pt]
WX Hyi & $ 0.23 ^{+0.07}_{-0.04} $ & $ 0.0748134(2) $ & $ 0.07737  $ &  & 0.0783 & $ 0.0342 _ { - 0 } ^ { + 0.0124 } $ & SU UMa & \useFN\shh$^,$\footnote{\nocite{SV81}{Schoembs} \& {Vogt} (1981)}$^,$\footnote{\nocite{B79}{Bailey} (1979)}$^,$\footnote{\nocite{WMF76}{Walker}, {Marino} \& {Freeth} (1976)} \\[3pt]
T Leo & $ 0.71(15) $ & $ 0.0588190(5) $ & $ 0.06022(2) $ & 0.06021 & 0.06025 & $ 0.0238 _ { - 0.0002 } ^ { + 0.0005 } $ & SU UMa IP? & \useFN\sh$^,$\footnote{\nocite{SSz84}{Shafter} \& {Szkody} (1984)}$^,$\footnote{\nocite{K97}{Kato} (1997)}\saveFN\kbc$^,$\footnote{\nocite{LPT93}{Lemm} {et~al.} (1993)}$^,$\useFN\kbc \\[3pt]
U Gem & $ 0.357(7) $ & $ 0.1769061898(30) $ & $ 0.20  $ & 0.197 & 0.203 & $ 0.131 _ { - 0.017 } ^ { + 0.017 } $ & U Gem(e) & \footnote{\nocite{NAL05}{Naylor}, {Allan} \& {Long} (2005)}$^,$\footnote{\nocite{Sm93}{Smak} (1993)}$^,$\footnote{\nocite{SW04}{Smak} \& {Waagen} (2004)}\saveFN\sw$^,$\useFN\sw$^,$\useFN\sw \\[3pt]
V603 Aql & $ 0.22(3) $ & $ 0.13809(12) $ & $ 0.14640(6)  $ & 0.144854 & 0.14686 & $ 0.0602 _ { - 0.0112 } ^ { + 0.0033 } $ & Fast nova & \footnote{\nocite{ACA00}{Arenas} {et~al.} (2000)}$^,$\footnote{\nocite{PTS93}{Patterson} {et~al.} (1993b)}\saveFN\pts$^,$ $^,$\footnote{\nocite{HM85}{Haefner} \& {Metz} (1985)}$^,$ \footnote{\nocite{PKS97}{Patterson} {et~al.} (1997)}\\[3pt]
UU Aqr & $ 0.30(7) $ & $ 0.163580429(5) $ & $ 0.17510(18) $ &  &  & $ 0.0704           $ & NL(E) & \footnote{\nocite{BSC94}{Baptista}, {Steiner} \&  {Cieslinski} (1994)}\saveFN\bsc$^,$\useFN\bsc$^,$\useFN\pkh \\[3pt]
DW UMa & $ 0.39(12) $ & $ 0.136606527(3) $ & $ 0.1454(1) $ &  & 0.1461 & $ 0.0644 _ { - 0 } ^ { + 0.0051 } $ & NL(E) & \footnote{\nocite{A-BKL03}{Araujo-Betancor} {et~al.} (2003)}$^,$\footnote{\nocite{SKB04}{Stanishev} {et~al.} (2004)}\saveFN\skbg$^,$\footnote{\nocite{PFT02}{Patterson} {et~al.} (2002b)}$^,$\useFN\skbg \\[3pt]
MV Lyr & $ 0.43 ^{+0.19}_{-0.13} $ & $ 0.132335  $ & $ 0.1377(4) $ &  & 0.1487 & $ 0.0405 _ { - 0 } ^ { + 0.0832 } $ & NL & \footnote{\nocite{HLS04}{Hoard} {et~al.} (2004)}$^,$\footnote{\nocite{RK03}{Ritter} \& {Kolb} (2003)}$^,$\footnote{\nocite{SPT95}{Skillman}, {Patterson} \&  {Thorstensen} (1995)}$^,$\footnote{\nocite{PS99}{Pavlenko} \& {Shugarov} (1999)} \\[3pt]
AM CVn & $ 0.18(1) $ & $ 0.011906623(3) $ & $ 0.012167 $ & $ 0.012164 $ & $ 0.012169 $ & $ 0.0218 _ { - 0.0002 } ^ { + 0.0002 } $ & AM CVn & \footnote{\nocite{RGN06}{Roelofs} {et~al.} (2006)}$^,$\footnote{\nocite{SPK99}{Skillman} {et~al.} (1999)}\saveFN\spk$^,$\useFN\spk$^,$\useFN\spk$^,$\useFN\spk \\[3pt]
KV UMa & $ 0.037(7) $ & $ 0.1699339(2) $ & $ 0.170529(6) $ & 0.17049 & 0.17073 & $ 0.0035 _ { - 0.0002 } ^ { + 0.0012 } $ & BHXRT & \footnote{\nocite{O01}{Orosz} (2001)}$^,$\footnote{\nocite{TCG04}{Torres} {et~al.} (2004)}$^,$\footnote{\nocite{UKM02}{Uemura} {et~al.} (2002a)}$^,$\footnote{\nocite{ZCS02}{Zurita} {et~al.} (2002)}$^,$\useFN\pkh \\
\multicolumn{2}{l}{(XTE J1118+480)} \\
QZ Vul & $ 0.042(12) $ & $ 0.3440915(9) $ & $  0.3469(1) $ &  & $ 0.3474 $ & $ 0.0082 _ { - 0} ^ { + 0.0014 }         $ & BHXRT & \footnote{\nocite{HHF96}{Harlaftis}, {Horne} \&  {Filippenko} (1996)}\saveFN\hhf$^,$\useFN\hhf$^,$\footnote{\nocite{CKP91}{Charles} {et~al.} (1991)}$^,$\footnote{\nocite{O'DC96}{O'Donoghue} \& {Charles} (1996)}\saveFN\odc \\
\multicolumn{2}{l}{(GS 2000+2)} \\
V1487 Aqr & $ 0.058(33) $ & $ 30.8(2) $ & $ 31.4 $ & 31.2 & 31.6 & $ 0.0195 _ { - 0.0065 } ^ { + 0.0065 } $ & BHXRT & \footnote{\nocite{HG04}{Harlaftis} \& {Greiner} (2004)}\saveFN\hg$^,$\footnote{\nocite{NBC06}{Neil} {et~al.} (2006)}\saveFN\nbc$^,$\useFN\nbc$^,$\useFN\nbc$^,$\useFN\nbc \\
\multicolumn{2}{l}{(GRS 1915+105)} \\[3pt]
V518 Per & $ 0.111 ^{+0.027}_{-0.033} $ & $ 0.2121600(2) $ & $ 0.2157(10) $ &  &  & $ 0.0167           $ & BHXRT & \footnote{\nocite{WNI00}{Webb} {et~al.} (2000)}\saveFN\wni$^,$\useFN\wni$^,$\footnote{\nocite{KMH95}{Kato}, {Mineshige} \& {Hirata} (1995)} \\
\multicolumn{2}{l}{(GRO J0422+32)} \\
GU Mus  & $ 0.13(2) $ & $ 0.432602(1) $ & $ 0.4376(10) $ &  &  & $ 0.0116           $ & BHXRT & \footnote{\nocite{OBM96}{Orosz} {et~al.} (1996)}$^,$\footnote{\nocite{CMC97}{Casares} {et~al.} (1997)}$^,$\useFN\odc \\
\multicolumn{2}{l}{(N Mus 1991)} \\
\hline
\end{tabular}

\setcounter{footnote}{\value{mpfootnote}}

\end{center}
\end{scriptsize}
\end{minipage}

\end{table*}

Our high-resolution 3D simulations, which are represented by blue squares in Figure~\ref{eq}, provide a far better match with observed systems than previous studies by \nocite{Mu98,Mu00}{Murray} (1998, 2000). As in \nocite{Mu00}{Murray} (2000) we see that simple dynamical precession as given by
\begin{equation} \label{eq:dyn}
\omega_{dyn}=p(r)\frac{q}{\sqrt{1+q}}\Omega_{\rm orb},
\end{equation}
poorly represents observed systems. This is true even if the location of the resonance for a gaseous disc, rather than for isolated particles, is used; the location of the resonance changes only by $\lesssim 1$ per cent. In a real gaseous disc the retrograde effect that pressure forces have on disc precession rates must be taken into account \nocite{L92,Mu00}({Lubow} 1992; {Murray} 2000). In a gaseous disc, the excited eccentricity propagates through the disc as a wave and is wrapped into a spiral by the differential precession of the gas. \nocite{Mu00}{Murray} (2000) assumed that the hydrodynamical precession is given by
\begin{equation} \label{eq:hyd}
\omega=\omega_{dyn}+\omega_{pr},
\end{equation}
where $\omega_{pr}$ is the pressure contribution to the precession, and showed that, under the assumption that the eccentricity is tightly wound (that is if it is wound up on a length-scale much smaller that the disc radius) then this pressure contribution at the 3:1 resonance radius is
\begin{equation} \label{eq:pr}
\omega_{pr}\simeq-\frac{2}{3}\Omega_{\rm orb}\left(\frac{c_{\rm s}}{\Omega_{\rm orb}\,a}\frac{1}{{\rm tan}\,i_{\rm p}}\right)^{2},
\end{equation}
where $i_{\rm p}$ is the pitch angle of the spiral wave, and $c_{\rm s}$ is the sound speed.

For each eclipsing SU\,UMa system we have calculated an inferred pressure contribution to the precession rate, $\omega_{\rm pr}$ (column 7 in Table~\ref{tb:eclipse}) in the manner of \nocite{Mu00}{Murray} (2000). A weighted mean gives $\omega_{\rm pr}=-1.36\,{\rm rad\,d^{-1}}$. This is combined with the mean mass of the WD in systems below the period gap \nocite{SD98}({Smith} \& {Dhillon} 1998) to calculate predicted hydrodynamical precession rates assuming that the precessional pressure contribution is similar for all systems.This is plotted as a solid curve on Figure~\ref{eq}. We note that \nocite{Pe06}{Pearson} (2006) considers the inclusion of pressure effects in an alternative way. Whilst a reasonable fit to the observations is achieved, the fact that there is a distribution in the eclipsing systems above and below this curve, demonstrates that the situation is not so simple. Possible contributory factors are distributions in primary mass or in disc temperature. We have plotted two further curves on Figure~\ref{eq}, one representing the hydrodynamical prediction for larger WD masses, and the other encompassing both larger $M_{1}$ and higher accretion disc temperature. These are the mean value of $M_{1}$ for the eight SU\,UMa eclipsing systems and the value of $\omega_{\rm pr}$ found by \nocite{Mu00}{Murray} (2000) respectively.
We tabulate $M_{1}$, the total system mass, $M_{\rm t}$, and the mid-plane sound speed at the resonance radius for each of the eclipsing SU\,UMa systems (Table~\ref{tb:eclipse}). These numbers do not seem to provide any clue as to the true cause of the spread in observed systems. In calculating $c_{\rm s}$, we assumed the same mass transfer rate throughout ($10^{-9}\,{\rm M_{\odot}\,yr^{-1}}$). Undoubtedly this, and consequently the temperature, varies from system to system.

Figure~\ref{eq} also shows the best fit that \nocite{GO06}{Goodchild} \& {Ogilvie} (2006) found to their analytic curve. They found that retrograde pressure terms are in fact only $\sim 1$ per cent of the dynamical term, far smaller than the values we infer above. They suggest that the offset of observation from dynamical precession rate be due instead to averaging over the disc, the eccentricity being distributed throughout the disc rather than being sharply peaked at the resonance itself. This is in accordance with our findings that dissipation-powered superhump overwhelmingly originates in the disc regions within $R_{3:1}$. \nocite{GO06}{Goodchild} \& {Ogilvie} (2006) found that their eccentricity distributions peaked close to $0.37\,a$, a value put forward by \nocite{P01}{Patterson} (2001) to match observation. We have plotted the dynamical curve as evaluated at $0.37\,a$ (dot-dashed light-blue line). We have also plotted two further lines which show dynamical curves evaluated at some fraction of the resonance radius, one to give the best weighted fit to the eclipsing SU\,UMa systems (red) and the second to give the best fit to our 3D simulations that have reached equilibrium, runs 5 to 10, (blue). We find values of $0.83\,R_{3:1}$ and $0.87\,R_{3:1}$ respectively.

\section{Discussion}

The precession rates for the simulated discs provide a much closer match with observation than has been achieved previously. There are a number of reasons for this. The improved mass resolution changes the stream--disc impact, leading to a different angular momentum distribution in the disc and a different precession rate. The radial sound speed distribution is correct for a steady-state disc (as opposed to an isothermal one as in the case of \nocite{Mu98}{Murray} (1998)), which means that the variation of density with radius is more realistic, which in turn will determine whether the disc precesses and at what rate. These updated simulations have hotter discs. This means the retrograde effect of pressure on precession will be greater \nocite{L91}({Lubow} 1991a). The viscosity is more in line with what is inferred for an $\alpha$-disc in the high state. The simulations of \nocite{Mu98}{Murray} (1998) were too viscous, which allowed the disc to grow and penetrate the resonance too easily. The propagation of the eccentricity inward through the disc would also have been inhibited. If the precession rate is dictated by a weighted average of the eccentricity as suggested by \nocite{GO06}{Goodchild} \& {Ogilvie} (2006), then the inhibition of inward eccentricity propagation would lead to high precession rates.
Finally, the extension to 3D changes the character of the resonance. We can see in Figure~\ref{compare} that the accretion disc look very different in 2D and 3D. One reason for this is that the character of the stream-disc interaction differs: for 2D mass injection all the particles follow one another exactly and so the stream tends to punch through the outer disc and deposit its angular momentum at $r < R_{3:1}$. Conversely in the 3D case the particles are more easily captured by the outer disc and so they effectively reduce the specific angular momentum of the outer disc. The growth and decline of the eccentric mode is strongly influenced by the interaction with the mass transfer stream. The picture is further complicated by the results of \nocite{KSH01}{Kunze}, {Speith} \& {Hessman} (2001) who found substantial stream overflow in their simulations. An important extension of this work would be to systematically isolate the importance of each of the effects which contribute to the disc precession rate.

It appears that the offset of both observation and simulation from dynamical expectations of precession rates is due to the averaging over the radii interior to $R_{3:1}$ which participate in the disc precession, and the inclusion of a retrograde effect of pressure forces is insufficient \nocite{GO06}({Goodchild} \& {Ogilvie} 2006). In Section~\ref{ecc} and Figure~\ref{discplots} we show that the eccentric instability is indeed manifest at radii as small as $0.15a$, so the majority of the disc area contributes to the mean precession rate. We note that \nocite{GO06}{Goodchild} \& {Ogilvie} (2006) required a very low semi-thickness parameter ($h=0.003$) to fit observations, whereas in our simulations this is an order of magnitude higher (Table~\ref{tb:discs}) and in better agreement with observational constraints on, and theoretical expectations of, disc thickness.

Empirically, systems which show superhumps generally have $q<0.24$. There are two classes of exceptions: nova-likes and U\,Gem, and magnetic systems. For the former, \nocite{O05}{Osaki} (2005) pointed out that Paczynski's derivation of $R_{\rm tides}$ assumes the disc to be cold. A sufficiently hot accretion disc may expand beyond this owing to a weakening of shocks by strong pressure effects at the last non-intersecting orbit. \nocite{O05}{Osaki} (2005) argues the persistently high-state nova-likes, and the unusually long 1985 outburst of U\,Gem in which superhumps were reported, may satisfy this temperature criterion. The magnetic systems, TV\,Col and possibly T\,Leo, may have discs which are pushed out by magnetic forces \nocite{RHA03}({Retter} {et~al.} 2003).

The high resolution 3D simulations presented here reproduce this distribution well; the upper limit for which superhumps are observed is also $q\simeq0.24$, albeit at this mass ratio it is a very slow process indeed. This also compares well with the theoretical upper limit of $q\lesssim0.25$. For $q=0.2422$ it appears that the eccentricity starts to grow more than once, but fails each time and the eccentricity falls away again, presumably because the encounter with the resonance was only marginal. The final eccentricity is highest for $q \sim 0.1$, and declines gradually as $q$ becomes less extreme. This is expected as the disc has more room to grow beyond the 3:1 resonance before being tidally truncated for extreme mass ratios. For the two most extreme mass ratio simulations, with $ q = 0.0526$ and $q = 0.0256$  the final eccentricity is negligible. In the $q=0.0256$  case there is no eccentricity growth. The $q=0.0526$ case is interesting: as the disc is growing, the eccentricity initially increases and superhumps are apparent, albeit weak and ill-formed (Figure~\ref{detail}). This eccentricity is, however, damped away again. What is the source of the damping? One possibility could be the action of the 2:1 resonance which may be excited in ultra-low mass ratio systems and which can act to damp eccentricity \nocite{L91b}({Lubow} 1991b). Perhaps this results in a close competition between the 2:1 and 3:1 resonance where the 2:1 resonance is marginally excited in the case of $q=0.0526$, and the 2:1 resonance only comes out on top once the disc has grown and this resonance is sufficiently populated. In reality, the most extreme mass ratio system in which superhumps are observed is the black hole X-ray transient KV\,UMa (XTE J1118 +480), which has $q=0.037\pm0.007$.
In 2D, the range for which the simulated discs become eccentric extends to much higher mass ratios ($q\leq0.4815$) (Table~\ref{tb:sph}, Figure~\ref{eq}). The confinement to 2D means that the character of the resonance is different and the strength of the resonance is increased.

We find that the growth rate of the eccentricity (the growth rate of $S_{(1,0)}$) is highly dependent on $q$, with high mass ratio systems taking a very long time indeed to become eccentric (Table~\ref{tb:sph}). We see in Figure~\ref{growth} that our new simulations have slower growth rates than previous 2D simulations by \nocite{Mu98}{Murray} (1998). Observationally, the rise times of superoutbursts is of order a day or so, and superhumps are generally detected within a day or so. The outburst onset is, however, probably governed by the thermal-viscous instability and hence its timescale is independent of the disc eccentricity. Generally there are no observations suitable for assessing whether or not the disc was eccentric before the superoutburst began, with intensive observations beginning after the rise to outburst, so timescales on which SU\,UMa discs typically develop their eccentricity is ill-constrained. We note further that our simulations do not include the thermal-viscous instability, so their evolution differs from that of SU\,UMa discs. \nocite{L91}{Lubow} (1991a) found an analytical expression for the growth rate of the eccentricity, finding it to be proportional to $q^2$. This is applied to an ideal narrow fluid ring. Clearly the eccentricity growth rate of our simulated discs is not proportional to $q^2$. \nocite{O05}{Osaki} (2005) uses the dependence of eccentricity growth on $q^2$ to propose, as a refinement to the TTI model, an explanation for type A/type B superoutbursts, namely those which show a precursor and those which do not. He suggests that it depends on whether the eccentricity growth rate is large enough to excite a significantly eccentric disc within the duration of a normal outburst, and that this is why most SU\,UMa systems having relatively low mass ratio show only Type B superoutbursts. However, if it is the case that high $q$ means lower growth rates as we see, then this argument fails, and probably other factors contribute.

\nocite{GO06}{Goodchild} \& {Ogilvie} (2006) also find extremely low growth rates in their analytical work. They use the growth term found by Lubow to describe the rate at which eccentricity is created at the resonance, but consider further how this eccentricity propagates through the disc. They explain their low growth rates as due to the eccentricity being strongly suppressed at the resonance itself. They find that the growth rates depend most strongly on mass ratio and on bulk viscosity, with further weaker dependence on disc semi-thickness. In column 7 of Table~\ref{tb:discs} we list the parameter, $h$, referred to by \nocite{GO06}{Goodchild} \& {Ogilvie} (2006) as the characteristic disc semi-thickness, and in column 11 the bulk viscosity at $R_{3:1}$ in our simulations. For the 3D simulations, these are $\sim0.036$ and $0.2$ respectively. Comparing our Figure~\ref{growth} with \nocite{GO06}{Goodchild} \& {Ogilvie} (2006)'s results, we see that the trend in our 3D points matches quite well with the $h=0.01$ line (the highest value of $h$ given) in their figure 10. In particular we see in both cases low growth rates at high $q$, a maximum at $q\sim0.08$ and a steep drop in growth rate at mass ratios below this. Our growth rates, though, are about a factor of 10 higher. Looking to their figure 9, this could be explained by our high semi-thickness parameter. Our bulk viscosity, though, is also rather high. It would be very interesting to make a study of empirical growth rates and their dependence on $q$ and on other known parameters to compare with these findings. This would require systematic monitoring of dwarf novae to catch the onset of outburst.

There are, however, distinct differences between the work of \nocite{GO06}{Goodchild} \& {Ogilvie} (2006) and the simulations that we present. Tidal modes, which would presumably act to truncate the eccentric mode, are not included in their work. The outer boundary conditions differ. We see too in Section~\ref{ecc} that the eccentricity distribution as a function of radius appears quite different from their findings. The situation in the simulations is further complicated by the presence and importance of the tidal 2-armed spiral structure which is not included in \nocite{GO06}{Goodchild} \& {Ogilvie} (2006)'s work.

In our simulations we see the superhump period decreasing (Figure~\ref{mass}) as is often observed over the course of a superoutburst \nocite{PBG93}({Patterson} {et~al.} 1993a). As the period changes, the disc eccentricity is increasing, and, in most cases, the disc mass has begun to decrease in response to the enhanced tidal torques. The superhump period decrease in the simulations can then be explained by the eccentric wave propagating inward, and additionally by radial shrinking of the disc. We are unsure how to explain the exception, $q=0.0526$, which shows an opposite behaviour for the initial part of the simulation. Perhaps it is related to the ideas of \nocite{UMI05}{Uemura} {et~al.} (2005), where they suggest an explanation for +ve $\dot{P}_{\rm sh}$ observed in a few cases. They suggest $\dot{P}_{\rm sh}$ is related to the amount of matter beyond $R_{3:1}$, so allowing for an outward propagation of the eccentric wave. For low $q$, the distance between $R_{3:1}$ and $R_{\rm tides}$ is greater.

The observational data points to a many-valued $\epsilon(q)$ relation (Figure~\ref{eq}). In particular 3 of the BHXRTs show systematically lower precession rates than those of CVs. QZ\,Vul (GS 2000+2) might be an exception to this. However the superhump period measurement is uncertain as it has been sparsely observed. The microquasar V1487 Aqr (GRS 1915+105) is a much longer period system and may not be comparable to the other BHXRTs. The accretion discs in BHXRTs are irradiated by the central X-ray source. We would expect these discs to be both hotter and also thicker due to the bloating effect of irradiation. Both of these would act to reduce the precession rate, due to the retrograde effect of pressure and due to the dependence on semi-thickness found by \nocite{GO06}{Goodchild} \& {Ogilvie} (2006). It would be very interesting if $\epsilon$ and $q$ for further BHXRTs could be determined. The only ultra-compact helium binary included, AM\,CVn, also lies below the main cluster of points in Figure~\ref{eq}. As \nocite{RGN06}{Roelofs} {et~al.} (2006) noted, the helium accretion disc in AM\,CVn could be thicker than its hydrogen-rich counterparts.

\section{Summary}

The main findings in this work can be summarised as follows:

\begin{itemize}
\item We present improved accretion disc simulations for a range of $q$. The main improvements are both numerical and physical: a higher mass resolution, extension to 3D, more realistic disc temperature and viscosity, and a radial dependence of sound speed appropriate to a steady-state accretion disc. We ran the simulations until equilibrium was reached. For $0.08 < q < 0.24$ the 3D discs reach an eccentric equilibrium and show a superhump signal in their energy dissipation rate (which we refer to as a simulated lightcurve).

\item The $\epsilon(q)$ dependence for the SPH simulations presented in this work shows a greatly improved match with observation than previous simulations.

\item
No high resolution 3D disc with $ q > 0.24$ developed superhumps. This agrees with theoretical expectations
and matches the majority of observations.

\item
The region of the disc within $R_{3:1}$ is overwhelmingly responsible for generating the dissipation-powered superhump in the simulations.

\item If the difference between observed precession rates and dynamical precession rates calculated at the 3:1 resonance radius is due to averaging over the disc as \nocite{GO06}{Goodchild} \& {Ogilvie} (2006) suggest, then we find that the best fit characteristic radius of the eccentricity distribution at which the dynamical precession rate is evaluated to be $0.87 R_{3:1}$ and $0.83 R_{3:1}$ for the 3D simulated discs and the observed eclipsing systems respectively. The differences between these two best-fit radii may be partly due to the differing surface density distributions in the two cases.

\item Our simulations show the effect of the increased efficiency of tidal return of angular momentum to the binary for an accretion disc which has become eccentric. The disc mass approaches a new lower steady-state value as the disc becomes eccentric. This is exactly as asserted by the TTI model. With the assumption of a radial dependence of viscosity, we deduce an effective $\sim 4$ per cent increase viscous torque between a disc which is circular and one that is eccentric. The increase depends on $q$.

\item As the eccentricity grows and the disc mass falls, the superhump period decreases.

\item The dependence of eccentricity growth rates on $q$ that we see in the simulations presented here is comparable to the work of \nocite{GO06}{Goodchild} \& {Ogilvie} (2006). Particularly, we find that for high mass ratios the growth rates are very low indeed, in contrast to the result of \nocite{L91}{Lubow} (1991a). This needs to be reconciled with observation.

\item
We show that superhumping discs have noticeable eccentricity even in their inner regions ($r \sim 0.15 a$). Conversely, non-superhumping discs are seen to be eccentric only in their outer regions. In this case however, this `eccentricity' is steady-state and has origin in tidal distortions, being therefore different from that which dominates the main body of the superhumping discs. We characterise the eccentricity distributions using two different methods.

\item
The disc motions can be described as superpositions of the S(k,l) modes, and the resonance radii act as nodes and antinodes in the complex standing wave dynamics the disc executes over a full precession period. We characterise the disc motions on $P_{\rm sh}$, the key timescale for the powering of the observed superhumps.

\item
The 4:1 and 5:1 resonances may  play roles in the dynamics of eccentric discs for $ q < 0.24 $. This may explain why the observed precession rates are closer to the dynamic precession rate at the 4:1 resonance than they are to the dynamic precession rate at the 3:1 resonance.

\item The observational data shows a multi-valued $\epsilon(q)$ relation. In particular, the BHXRTs show systematically lower precession rates than those of the CVs, which may be expected when the higher temperature and thickness of their irradiated discs is considered.

\end{itemize}

\section{Acknowledgements}

Helpful comments from the referee were much appreciated. We acknowledge the use of the supercomputing facilities at the Centre for Astrophysics and Supercomputing, Swinburne University of Technology. AJS was supported by a PPARC studentship.


\label{lastpage}

\end{document}